\documentclass[twocolumn,a4paper]{article}
\usepackage[utf8]{inputenc}
\usepackage[T1]{fontenc}
\usepackage{amsmath,amssymb,amsfonts}
\usepackage{graphicx}
\usepackage{xcolor}
\usepackage{hyperref}
\usepackage{cite}
\usepackage{geometry}
\usepackage{abstract}
\usepackage{graphicx}
\usepackage{array}

\title{\vspace{-1.5em}\textbf{Beyond the Bounce: Multiple Tidal Sign Reversals and Turning-Point Bifurcations in Multi-Horizon Black Holes}}
\author{\textbf{Mohammad Ali S. Afshar}\textsuperscript{1,2}\thanks{m.a.s.afshar@gmail.com}, \textbf{J. Sadeghi}\textsuperscript{1}\thanks{pouriya@ipm.ir}. \\[0.5em]
\small\textsuperscript{1}Department of Theoretical Physics, Faculty of Basic Sciences, University of Mazandaran, P. O. Box 47416-95447, Babolsar, Iran \\
\small\textsuperscript{2}School of Physics, Damghan University, P. O. Box 3671641167, Damghan, Iran \\
}

\newcommand{\keywords}[1]{\par\vspace{0.3em}\noindent\textbf{Keywords:} #1}
\begin{document}
\maketitle
\thispagestyle{empty}

\begin{abstract}
We investigate the radial motion and tidal forces experienced by neutral test particles in multi-horizon black hole solutions arising from Einstein gravity coupled to nonlinear electrodynamics (NED). Focusing on the three- and four-horizon configurations, we examine how nonlinear electromagnetic corrections modify the causal structure, radial geodesic motion, and tidal-force profiles in comparison with the Schwarzschild and Reissner-Nordström (R-N) spacetimes.

Our analysis shows that the NED field gives rise to multiple zero crossings in both the radial and angular tidal-force components, leading to successive transitions between stretching and compressive tidal regimes. More importantly, the radial equation of motion contains classically forbidden regions bounded by bounce-back points. For the class of trajectories considered in this work, these forbidden regions prevent particles from entering the spacetime domain where the tidal forces become divergent. In the super-extremal regime admitted by these solutions, the forbidden region may extend beyond the event horizon, preventing particles released from rest at sufficiently large distances from crossing the horizon.

We further identify a systematic ordering of the critical charge values associated with the appearance of tidal-force zero crossings, additional horizons, and bounce-back points. For both the three- and four-horizon configurations, these critical values satisfy a hierarchical ordering, indicating that changes in the tidal-force structure precede the corresponding modifications of the horizon configuration.

These results demonstrate that nonlinear electrodynamics can substantially modify the classical dynamics of neutral particles in multi-horizon black hole spacetimes through the combined effects of forbidden regions, multiple tidal transitions, and changes in the horizon structure.
\end{abstract}

\keywords{Tidal Force, Multi-Horizon Black Holes, Geodesic Deviation, Bounce-Back points }

\section{Introduction}
In Newtonian gravity, tidal forces  are defined as differential accelerations between nearby  freely falling particles and have long been studied in celestial mechanics. Notable examples include  tidal locking and tidal heating, where variations in tidal  forces along eccentric orbits lead to internal dissipation  of mechanical energy ~\cite{1,2}. Consequently, the encounter with these forces at deeper layers of physical understanding was far from unexpected. In General Relativity, as our purely Newtonian physical interpretation of gravity was fundamentally absorbed  into the geometric curvature of spacetime, tidal effects were naturally interpreted as manifestations  of spacetime curvature through the Riemann tensor acting on geodesic deviation.  These effects could be rigorously analyzed through the   geodesic deviation and the study of Jacobi  fields ~\cite{3,4,5,6}.\\
The emergence of black holes as one of the most striking predictions of General Relativity provided a pristine and exceptional arena for investigating this class of forces. The strong curvature near black holes significantly enhances tidal effects on nearby matter. Unlike the conventional gravitational interaction between point masses, the tidal effect- arising from the gradient of the gravitational field- causes the side of a continuous body closer to the gravitational source to experience a greater acceleration than its far side. This leads to radial stretching and angular compression  ~\cite{7}. In the simplest case, in Schwarzschild spacetime, infalling bodies experience  radial stretching and angular compression, leading to the  well-known spaghettification effect  ~\cite{7,8}.\\
Gradually, as our theoretical understanding of black holes deepened and new physical parameters were incorporated into black hole geometries, our comprehension of tidal behavior was dramatically transformed. In the R–N black hole, electric  charge modifies tidal forces and can lead to sign changes  in radial and angular components depending on the radial position ~\cite{9}. This sign change, occurs because of the adding charge contribution to the metric function and  altering the relative dominance of its contribution at different radii and consequently  affecting the derivatives $f'(r)$ and $f''(r)$ that govern tidal forces. This unexpected behavior has been investigated in other static geometries, including Kiselev black holes, regular black holes (such as the Bardeen model), and even naked singularities ~\cite{10,11,12,13,14,15,16,17,18,19,20,21,22,23,24,25,26,27,28,29,30}. Furthermore, the analysis of tidal forces in the rotating Kerr spacetime has revealed additional complexities associated with the ergosphere region ~\cite{31,32}.\\
Today, it is well established that the study of tidal forces is by no means a purely theoretical endeavor; rather, it constitutes a key to unraveling the mysteries of high-energy cosmic phenomena. When a star approaches a supermassive black hole sufficiently closely, tidal forces tear it apart, giving rise to an event known as a tidal disruption event (TDE) ~\cite{33}. These events, by producing luminous emissions across the X-ray, ultraviolet, and optical bands, provide invaluable information about the mass and spin of the central black hole ~\cite{34,35}. Beyond this, Wheeler proposed that tidal interactions with stars in the ergosphere of rotating black holes could serve as a mechanism for the production of relativistic jets ~\cite{36}. More recently, the utilization of tidal acceleration within gravitational wave detectors such as LIGO has been proposed as a novel method for the direct measurement of graviton effects ~\cite{37,38,39,40,41}. The significance of this subject is further underscored by the fact that the disruption boundary- the Roche limit- is mass-dependent: for stellar-mass black holes, tidal disruption occurs well before the event horizon is reached, whereas for supermassive counterparts, a body may traverse the horizon intact ~\cite{42,43}. It is evident that each alternative compact object model (including regular black holes, naked singularities, or wormholes) will possess its own distinctive tidal signature, and comparative studies of these signatures lay the groundwork for discriminating among such models in future observations ~\cite{44,45,46,47,48,49,50}.\\\\
As discussed above, the incorporation of parameters arising from fields coupled to gravity in the black hole action leads to substantial modifications in the tidal force structure compared to the simple Schwarzschild model. Among the various types of fields, electromagnetic fields have undoubtedly been the most extensively studied option. Initially, the combination of General Relativity with Maxwell electrodynamics appeared to be a successful descriptive framework for studying the universe. However, our progressively deepening understanding revealed that both theories require corrections in extreme regimes (ultra-strong fields or when quantum effects become significant). In this context, Nonlinear Electrodynamics, as a controlled generalization of Maxwell theory that often emerges from effective quantum theories, provides an ideal framework for exploring these corrections. When NED fields are coupled to gravity, they give rise to charged black hole solutions that deviate from the RN geometry. These deviations can not only improve the interior behavior of spacetime and lead to the construction of regular black holes but also give rise to more complex causal structures, including configurations with multiple horizons. For this reason, NED has become a prevalent tool in the modeling of novel black holes and the investigation of their observational properties ~\cite{51,52,53,54,55,56,57,58,59,60,61,62,63,64,65,66}.\\
 In line with these efforts, Gao et al. recently succeeded in finding charged black hole solutions with more than two horizons (multi-horizon) within the framework of Einstein-Nonlinear Electrodynamics theory, and specifically provided explicit expressions for the NED Lagrangian corresponding to multi-horizon black holes ~\cite{67}. These multi-horizon black holes exhibit richer physics, including phenomena such as evaporation and antievaporation occurring at inner horizons. These phenomena are particularly relevant in the context of quantum effects near Cauchy-like horizons and may have implications for the stability of the internal structure of such spacetimes.  Although the geometric, thermodynamic, and orbital properties of multi-horizon black holes in the NED framework have been extensively studied, our understanding of the dynamics of infalling bodies and, in particular, the behavior of tidal forces in these complex spacetimes remains incomplete. Given that tidal forces are highly sensitive to the derivatives of the metric function, the alteration of the horizon structure and the presence of nonlinear terms in the electromagnetic Lagrangian will inevitably have profound effects on the tidal profile and geodesic deviation. In this paper, we present a comprehensive investigation of tidal forces and geodesic deviation for radially freely falling bodies in the background of a multi-horizon black hole within NED theory. Our primary objective is to uncover how the multi-horizon structure and nonlinear parameters influence both the qualitative and quantitative modifications of tidal forces.

\section{Key Concepts and Definitions}
To  analyze tidal dynamics in multi-horizon black hole spacetimes within nonlinear electrodynamics (NED), we first  introduce the formal framework based on geodesic deviation. We consider the  static, spherically symmetric line element as:
\begin{equation}
    ds^2 = -\mathfrak{f}(r)dt^2 + \frac{1}{\mathfrak{f}(r)}dr^2 + r^2 (d\theta^2 + \sin^2\theta d\phi^2),
    \label{eq:metric}
\end{equation}
where $\mathfrak{f(r)}$ determines the causal and horizon structure of the spacetime. 
\subsection{Geodesic Deviation and the Tidal Tensor}
The relative acceleration  between nearby freely falling particles is governed by the  geodesic deviation equation. For physical interpretation, we project the  equation onto a locally orthonormal tetrad frame comoving with a radially infalling observer. The tetrad basis vectors $e_{\hat{a}}^{\ \mu}$, satisfy $g_{\mu\nu}e_{\hat{a}}^{\ \mu}e_{\hat{b}}^{\ \nu} = \xi_{\hat{a}\hat{b}}$, with $e_{\hat{0}}^{\ \mu}=u^\mu$ chosen as the observer’s four-velocity. 

In this local frame, the geodesic deviation equation takes the form:
\begin{equation}
    \frac{D^2 \xi^{\hat{a}}}{d\tau^2} = - R^{\hat{a}}_{\ \hat{0}\hat{b}\hat{0}} \xi^{\hat{b}} \equiv \mathcal{T}^{\hat{a}}_{\ \hat{b}} \xi^{\hat{b}},
    \label{eq:deviation}
\end{equation}
here $\xi^{\hat{a}}$ is the Jacobi field describing spatial separation between nearby geodesics, $\tau$ is the proper time, and $R^{\hat{a}}_{\ \hat{0}\hat{b}\hat{0}}$ are the components of the Riemann curvature tensor projected onto the tetrad basis. The matrix $\mathcal{T}^{\hat{a}}_{\ \hat{b}}$  is the \textit{tidal tensor};  its eigenvalues determine the stretching or compressing  character of tidal forces.

To fix conventions, we explicitly construct an orthonormal tetrad basis $e_{\hat{a}}^{\ \mu}$ adapted to a radially freely falling observer. For the metric (\ref{eq:metric}) with the signature $(-,+,+,+)$, we choose
the following tetrad components as:
\begin{align}
    e_{\hat{0}}^{\ \mu} &= \left( \frac {\mathcal{E}}{\mathfrak{f}(r)}, -\sqrt{\mathcal{E}^2 - \mathfrak{f}(r)}, 0, 0 \right), \\
    e_{\hat{1}}^{\ \mu} &= \left( -\frac{\sqrt{\mathcal{E}^2 - \mathfrak{f}(r)}}{\mathfrak{f}(r)}, \mathcal{E}, 0, 0 \right), \\
    e_{\hat{2}}^{\ \mu} &= \left( 0, 0, \frac{1}{r}, 0 \right), \\
    e_{\hat{3}}^{\ \mu} &= \left( 0, 0, 0, \frac{1}{r \sin\theta} \right),
\end{align}
where $\mathcal{E} = \sqrt{\mathfrak{f}(b)}$ is the conserved specific energy for a particle released from rest at $r=b$. It is straightforward to verify that these vectors satisfy the orthonormality condition $g_{\mu\nu}e_{\hat{a}}^{\ \mu}e_{\hat{b}}^{\ \nu} = \xi_{\hat{a}\hat{b}}$, with $\xi_{\hat{a}\hat{b}} = \text{diag}(-1, 1, 1, 1)$. Projecting the Riemann tensor onto this specific tetrad basis yields the exact diagonal components of the tidal tensor $\mathcal{T}^{\hat{a}}_{\ \hat{b}}$ used in the subsequent analysis.

\subsection{Radial and Angular Tidal Forces}
For the metric (\ref{eq:metric}), spherical symmetry  ensures that the tidal tensor is diagonal in the chosen  frame. The non-vanishing independent components yield the radial and angular tidal forces. Specifically, the radial and the angular tidal forces ($\eta_{\text{radial}}$,$\eta_{\text{angular}}$) are given by the corresponding eigenvalues of the tidal tensor:
\begin{align}
    \eta_{\text{radial}} &\equiv \mathcal{T}^{\hat{1}}_{\ \hat{1}} = -\frac{1}{2} \mathfrak{f''}(r), \label{eq:rad_force} \\
    \eta_{\text{angular}} &\equiv \mathcal{T}^{\hat{2}}_{\ \hat{2}} = \mathcal{T}^{\hat{3}}_{\ \hat{3}} = -\frac{\mathfrak{f'}(r)}{2r}, \label{eq:ang_force}
\end{align}
where primes denote derivatives with respect to the radial coordinate $r$. 
A positive value  corresponds to radial stretching, while a negative value  corresponds to compression. In the standard Schwarzschild spacetime ($\mathfrak{f}(r) = 1 - 2M/r$), $\eta_{\text{radial}}$ is strictly positive and $\eta_{\text{angular}}$ is strictly negative everywhere outside the singularity, leading to the ubiquitous 'spaghettification' effect. Charge and nonlinear electromagnetic  corrections modify $\mathfrak{f}(r)$, which can lead to sign changes in tidal-force components.

\subsection{Kinematic Constraints and Turning Points}
To determine the trajectory of a radially infalling test particle, we utilize the conservation of specific energy $\mathcal{E}$, associated with the timelike Killing vector $\xi^\mu = (\partial_t)^\mu$. For a particle released from rest at a radial coordinate $r = b$, the conserved energy is $\mathcal{E} = \sqrt{\mathfrak{f}(b)}$. The radial equation of motion becomes:
\begin{equation}
    \left(\frac{dr}{d\tau}\right)^2 = \mathcal{E}^2 - \mathfrak{f}(r) = \mathfrak{f}(b) - \mathfrak{f}(r).
    \label{eq:radial_velocity}
\end{equation}
A \textit{turning point} (or stopping radius), $R_{\text{stop}}$, occurs when the radial velocity vanishes, i.e., $dr/d\tau = 0$. From Eq. (\ref{eq:radial_velocity}), this implies:
\begin{equation}
    \mathfrak{f}(R_{\text{stop}}) = \mathfrak{f}(b).
    \label{eq:turning_point}
\end{equation}
In uncharged spacetimes, the monotonic nature of $\mathfrak{f}(r)$ ensures that $R_{\text{stop}} \to 0$ as the particle approaches the singularity. Conversely, in charged and  NED spacetimes, non-monotonic behavior of $\mathfrak{f}(r)$ can  produce additional real roots corresponding to turning  points inside the geometry. The existence of such turning points implies that, mathematically, the particle experiences a repulsive core and undergoes a radial 'bounce' before reaching the central singularity, a feature that profoundly alters the internal causal structure and the maximal analytic extension of the spacetime.

\subsection{Critical Radii: Zeros and Extrema of Tidal Forces}
A key aspect of tidal dynamics in modified spacetimes is the identification of critical radii where the tidal forces exhibit qualitative transitions ( change sign or reach extrema). We  distinguish two types of critical radii:\\
\textbf{1. Zero-crossing Radii (Sign Inversions):}\\
These radii, $R_{\text{zero}}$, mark the boundaries where the tidal force indicating a change of sign and  transitions from stretching to compression, or vice versa.  
They are determined by setting the tidal force components to zero:
\begin{equation}
\begin{aligned}
    \text{Radial:} \quad & \mathfrak{f}''(R_{\text{zero}}^{\text{rad}}) = 0, \\
    \text{Angular:} \quad & \mathfrak{f}'(R_{\text{zero}}^{\text{ang}}) = 0.
\end{aligned}
\label{eq:zero_crossing}
\end{equation}

\textbf{2. Extremum Radii (Maxima and Minima):}
These are radii, $R_{\text{ext}}$,  where tidal-force magnitudes attain  local extrema, indicating maximal tidal stress on an infalling body and could be found by differentiating the force components with respect to $r$ and setting the derivatives to zero:
\begin{equation}
\begin{aligned}
\text{Radial Extremum:} \quad & \frac{d}{dr}\left[ \frac{1}{2}\mathfrak{f}''(r) \right]_{r=R_{\text{ext}}^{\text{rad}}} = 0 \\
\Rightarrow \mathfrak{f}'''(R_{\text{ext}}^{\text{rad}}) = 0\\
\text{Angular Extremum:} \quad & \frac{d}{dr}\left[ \frac{\mathfrak{f}'(r)}{2r} \right]_{r=R_{\text{ext}}^{\text{ang}}} = 0\\
&\Rightarrow R_{\text{ext}}^{\text{ang}} \mathfrak{f}''(R_{\text{ext}}^{\text{ang}})- \mathfrak{f}'(R_{\text{ext}}^{\text{ang}}) = 0.
\end{aligned}
\label{eq:extremum}
\end{equation}
\section{The Multi-Horizon Black Hole Model in Nonlinear Electrodynamics}
For the investigation of tidal dynamics in the presence of strong electromagnetic fields, we employ the framework of Einstein  gravity coupled to a generalized nonlinear electrodynamics (NED) sector, as formulated by Gao ~\cite{67}. This model admits exact  analytical black hole solutions with more than two horizons, leading to a geometric structure that differs substantially from that of the standard R–N spacetime.\\
\subsection{Action and Field Equations}
The action for this model is given by:
\begin{equation}
    S = \int d^4x \sqrt{-g} \left( R + \mathcal{L}_{EM} \right),
    \label{eq:action}
\end{equation}
where $R$ is the Ricci scalar and $\mathcal{L}_{EM}$ is the generalized electromagnetic Lagrangian. To encompass a broad class of NED theories, including the Euler–Heisenberg and Born–Infeld models, the electromagnetic Lagrangian is expressed as an infinite power series in the electromagnetic invariant $F^2 \equiv F_{\mu\nu}F^{\mu\nu}$:
\begin{equation}
    \mathcal{L}_{EM} = \sum_{i=1}^{\infty} \alpha_i (F^2)^i,
    \label{eq:lagrangian}
\end{equation}
where $\alpha_i$ are dimensional coupling constants with the dimension of $[\text{length}]^{2(i-1)}$. In the limit $\alpha_1 = 1$ and $\alpha_{i>1} = 0$, the theory reduces smoothly to standard Einstein–Maxwell electrodynamics.
Assuming a static, spherically symmetric spacetime, the line element and the electromagnetic take the form:
\begin{align}
    ds^2 &= -\mathfrak{f}(r)dt^2 + \frac{1}{\mathfrak{f}(r)}dr^2 + r^2 (d\theta^2 + \sin^2\theta d\phi^2), \label{eq:metric_ansatz} \\
    A_\mu &= [\Phi(r), 0, 0, 0].
\end{align}
Substituting this  ansatz into the Einstein and generalized Maxwell equations yields the metric function $\mathfrak{f}(r)$ and the electric potential $\Phi(r)$ as infinite series in inverse powers of $r$. 
 \subsection{Structural and Physical Properties}
Before analyzing the tidal  forces, we summarize the structural and physical properties that distinguish these NED multi-horizon black holes  from the standard RN solution and from conventional  regular black hole models  ~\cite{67}:
\begin{enumerate}
\item \textbf{Parameter Independence of the Lagrangian:}\\ Unlike many regular black hole models in the literature (e.g., the Bardeen or Ayon-Beato-Garcia solutions), in  which the electromagnetic Lagrangian explicitly depends on the macroscopic parameters $M$ and $q$, the NED Lagrangians derived in this framework depend \textit{solely} on the intrinsic coupling constants $\alpha_i$. This represents a significant theoretical advantage, as the nonlinear electromagnetic sector remains decoupled from the global thermodynamic parameters of the black hole 
\item \textbf{Super-Extremal Charge Configurations:}\\ In the standard RN spacetime, the condition $q > M$ eliminates the event horizon, leading to a naked singularity. In the present NED multi-horizon models,  however, the higher-order inverse-power corrections  modify the interior geometry such that event horizons may still exist for charge-to-mass ratios exceeding those allowed in the RN solution. Within  the corresponding parameter range, the central curvature singularity remains hidden behind an event  horizon.
\item \textbf{Additional Hair Parameters:}\\ A black hole solution possessing $N+1$ horizons is characterized by $N+1$ independent physical parameters: the mass $M$, the electric charge $q$, and $N-1$ nonlinear coupling constants $\alpha_i$. In this sense, these multi-horizon black holes  possess additional “hair” arising from the nonlinear  electromagnetic sector, which modifies the causal  structure, the effective potential governing geodesic  motion, and the thermodynamic properties of the  solution.
\item \textbf{Vanishing Tidal Love Numbers (TLNs):}\\ A remarkable feature of this model is its static response to external perturbations. For  the three-horizon black hole solution, the tidal Love numbers, which quantify the static tidal  deformability of a compact object, vanish exactly. Consequently, the three-horizon NED black hole  shares the same vanishing static tidal response as  the Schwarzschild black hole. Physically, this implies that despite the complex internal layering and higher-order electromagnetic corrections, the event horizon exhibits no static deformability in response to an external tidal field. The vanishing TLNs constrain the external multipolar response, leaving the  asymptotic static tidal behavior indistinguishable  from that of classical vacuum black holes. 
\item \textbf{Orbital Dynamics and the Penrose Process:}\\ The multi-horizon structure significantly alters the geodesic motion. The model admits stable circular orbits for both null and timelike geodesics. Notably, the Innermost Stable Circular Orbit (ISCO) for null geodesics is shifted  outward ($\epsilon \simeq 1.5055$) relative to the Schwarzschild  solution, whereas the timelike ISCO is shifted closer  to the event horizon ($\epsilon \simeq 2.0678$). Furthermore,  the Penrose process can occur even when the test  particle and the black hole possess charges of the  same sign, provided that the coupling constant $\alpha_2$  is sufficiently negative.
 \item \textbf{Rich Internal Causal Structure:}\\ The presence of multiple horizons gives rise to an internal causal structure that is absent in standard one- or two-horizon  spacetimes. As the  charge $q $ and the coupling parameters are varied,  two horizons may approach a degenerate configuration corresponding to a Nariai-type limit. Near such extremal configurations, quantum backreaction effects may become important  and can potentially lead to phenomena such as antievaporation and mass-inflation instabilities near  the inner Cauchy-like horizons. These features suggest that  the internal causal structure is sensitive to both  classical perturbations and quantum corrections. 
 \end{enumerate}
 \section{ Tidal Forces and the III-Horizon structure}
 Considering the general structure of the metric presented in ~\cite{67} and setting $\alpha_1 = 1$, $\alpha_2 \neq 0$,  together with the truncation relations $\alpha_3 = 4\alpha_2^2$, $\alpha_4 = 24\alpha_2^3$, and so forth, the metric function and electric potential for the three-horizon configuration are given by:
\begin{align}
    \mathfrak{f}(r) &= 1 - \frac{2M}{r} + \frac{q^2}{r^2} + \frac{2\alpha_2 q^4}{5 r^6}, \label{eq:U_3horizon} \\
    \Phi(r) &= \frac{q}{r} + \frac{4\alpha_2 q^3}{5 r^5},
\end{align}
where $M$ and $q$ denote the mass and electric charge of the black hole, respectively. For  sufficiently negative values of $\alpha_2$ and appropriate parameter choices, the equation $\mathfrak{f}(r) = 0$ admits three distinct positive real roots, corresponding to two inner Cauchy-like horizons and one outer event horizon. 
Because of the complexity  of the metric function, a purely analytical investigation of  the multi-horizon configurations is generally not feasible.  Accordingly, we rely on numerical analysis throughout  this paper to investigate the properties and tidal dynamics of the spacetime. Furthermore, since  our primary objective is to investigate the influence of  nonlinear electrodynamics on tidal-force dynamics, and  because the physically admissible values of the coupling  parameter $\alpha$ are typically small, we focus primarily on  the effects of varying the black hole charge. This methodological choice revealed an interesting feature. Our numerical results indicate that  the spacetime exhibits three distinct behavioral regimes,  determined primarily by the magnitude of the black hole  charge. One of  the main findings of this work is the sequence in which  these regimes emerge as the black hole charge varies, and  this behavior will be discussed in detail in the following  sections. 
 \subsection{Horizons Emergence  Behavioral Regime}
\begin{figure}[htbp]
    \centering
    \includegraphics[width=1.01\linewidth,height=6cm]{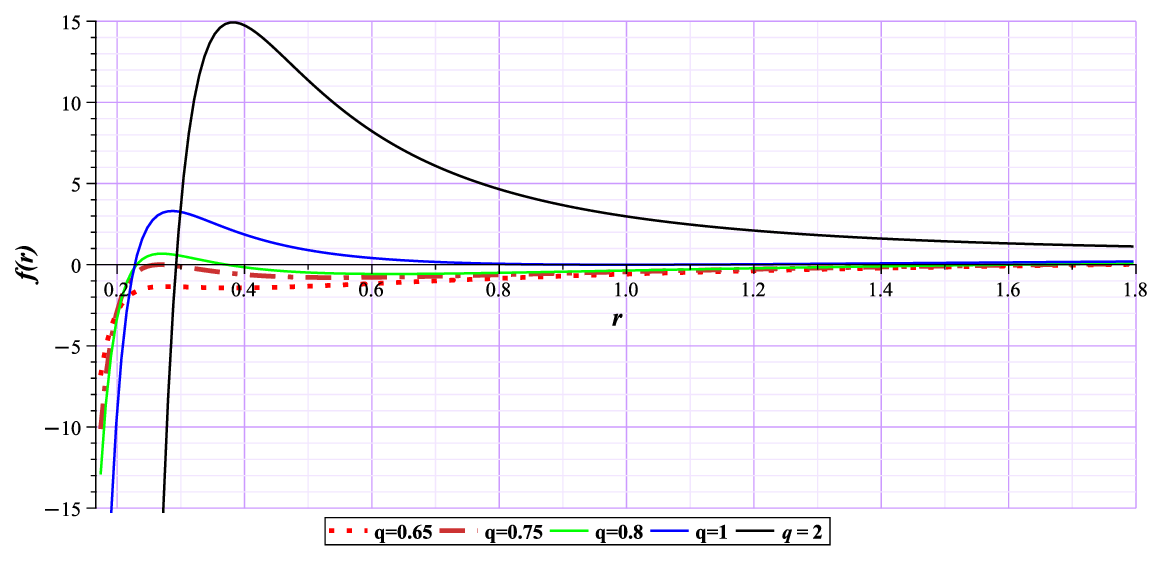}
    \caption{Fig. 1: Metric function with $\alpha_2 = -0.004$ and $M = 1$ and different $q$ for III-Horizon structure of  NED Black hole.}
    \label{1}
\end{figure}

Fixing the representative values $\alpha_2 = -0.004$ and $M = 1$, Fig.~\ref{1}shows that the black hole possesses a single-horizon configuration until $q = 0.749$. Once this  critical value is exceeded, two new horizons emerge. Following the standard terminology, we identify  the outermost horizon as the event horizon and the innermost horizon as the Cauchy horizon.
It is worth noting here that in standard black hole models, as the system approaches the extremal limit, the event and Cauchy horizons gradually converge, ultimately coinciding at extremality. In contrast, for the present model, the innermost horizon remains nearly stationary over the considered parameter range, whereas the intermediate horizon  moves toward the event horizon and eventually merges  with it. Consequently, the intermediate horizon plays  the role usually associated with the Cauchy horizon in the approach to extremality. Furthermore, as reported  in ~\cite{67} and summarized in Sec. 2.2, , this framework appears to accommodate super-extremal charges up to significantly large ratios, $q/m \gg 1$ . In this regime, the two outer horizons disappear, while the remaining inner horizon moves outward  toward larger values of $r$. This evolutionary behavior is clearly depicted in Fig.~\ref{1} for a charge of $q = 2$.
 \subsection{ $R_{\text{stop}}$ Emergence  Behavioral Regime} 
\begin{figure}[htbp]
    \centering
    \includegraphics[width=1.01\linewidth, height=6.5cm, ]{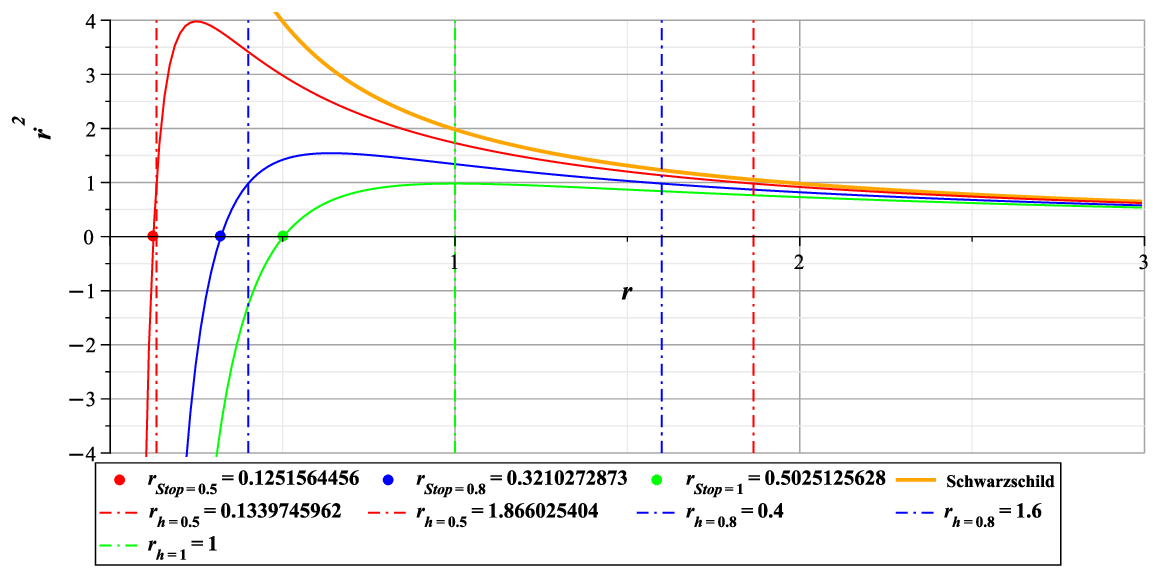}
    \caption{ $r^\cdot{^2}$ function with $M = 1$ and different $q$ for R-N and Schwarzschild Black hole.}
    \label{2}
\end{figure}

\begin{figure}[htbp]
    \centering
    \includegraphics[width=1.01\linewidth]{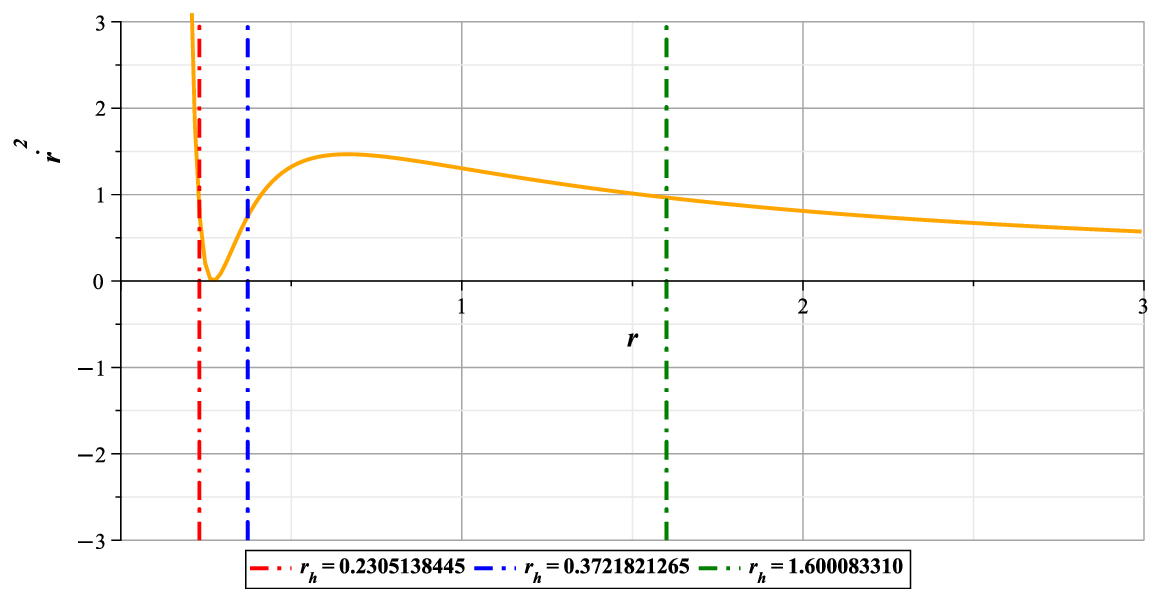}
    \caption{ $r^\cdot{^2}$  function with $\alpha_{2} = -0.004, b = 100, M = 1, q = 0.8222$ for III-Horizon structure of  NED Black hole.}
    \label{3}
\end{figure}

Previous investigations and the analysis of (\ref{eq:radial_velocity}) for the R-N metric structure reveal the existence of a bounce-back point at $r_{\text{Stop}} = \frac{bq^2}{2Mb - q^2}$ in the dynamical behavior of test particles. As shown in Fig.~\ref{2}, the  sign of $\dot r^2$ changes across this point, indicating that the  particle reverses its radial motion there. Strictly speaking, the initial position of the particle at $r=b$, from which it starts its radial fall from rest, is also a root of Eq.~(\ref{eq:radial_velocity}). Throughout this paper, however, we reserve the term \emph{bounce-back point} exclusively for the nontrivial turning point encountered during the subsequent motion and therefore do not refer to the initial position as a bounce-back point.
Within the maximal analytic extension of the RN spacetime, the particle would mathematically continue into another asymptotically flat region after reaching $R_{\mathrm{stop}} $~\cite{9}. However, once the physical instability of the Cauchy horizon is taken into account, together with the fact that $R_{\mathrm{stop}}$ always lies inside the inner horizon, this portion of the maximal analytic extension is generally regarded as physically inaccessible~\cite{9}.\\ 
Nevertheless, this characteristic feature of the R-N solution illustrates the influence of the Maxwell field on particle dynamics and contrasts with the purely gravitational Schwarzschild spacetime. The Schwarzschild solution possesses no such turning point. Instead, Eq.~(\ref{eq:radial_velocity}) yields $\dot r^2>0$ throughout the entire trajectory, as illustrated by the orange curve in Fig.~\ref{2}.\\
Based on these  results, when we examine our present model incorporating NED, we discover that its behavior is substantially more intricate  than that of conventional models.  Specifically, for charge values up to the threshold $q=0.8222$, the function $\dot r^2$ remains positive despite exhibiting non-monotonic deviations from the Schwarzschild profile, still exhibit Schwarzschild-like behavior. Consequently, no bounce-back point is present within this parameter range (Fig.~\ref{3}).\\
A comparison with the horizon-emergence regime shows that, although the NED corrections generate an additional intermediate horizon for $q\gtrsim0.75$, the radial dynamics remain qualitatively Schwarzschild-like over this interval, indicating that the gravitational contribution still dominates the particle motion. The appearance of nonzero extrema nevertheless signals the gradual influence of the nonlinear electromagnetic field on the radial dynamics.\\ 
\begin{figure}[htbp]
    \centering
    \includegraphics[width=1.01\linewidth]{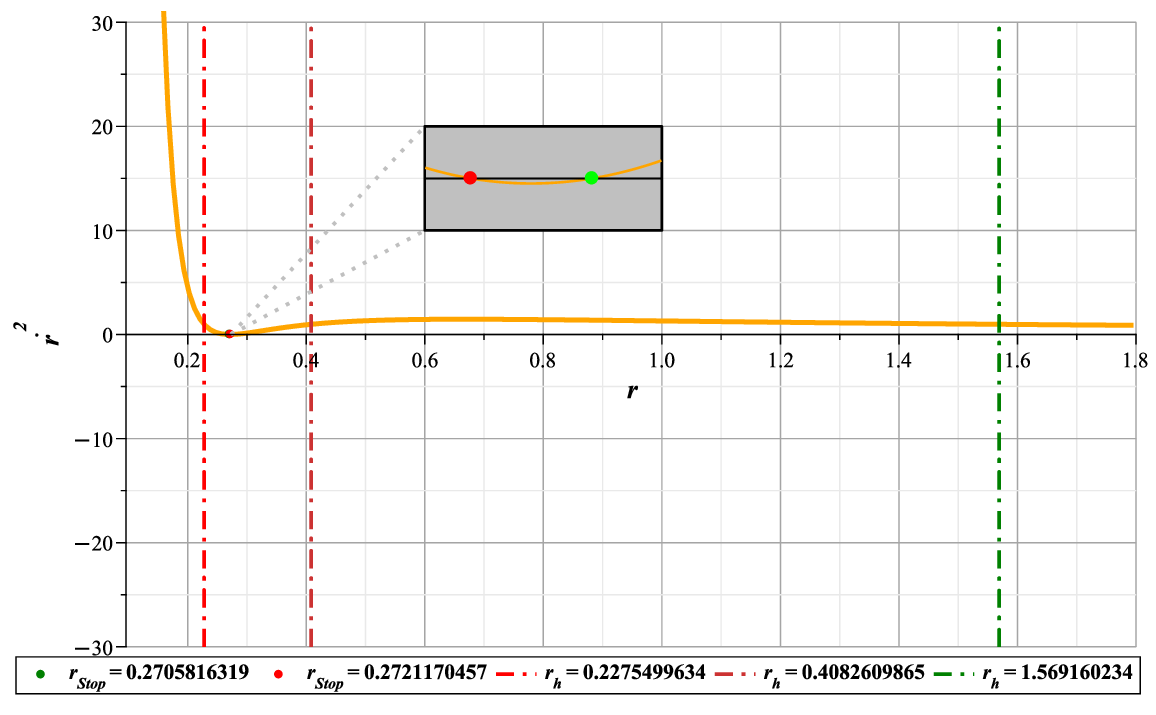}
    \caption{ $r^\cdot{^2}$  function with $\alpha_{2} = -0.004, b = 100, M = 1, q = 0.8223$ for III-Horizon structure of  NED Black hole.}
    \label{4}
\end{figure}
When the charge reaches the critical value $q=0.8223$, a qualitatively RN-like radial behavior emerges through the appearance of bounce-back points. However, a highly intriguing feature is that instead of a single turning point, the system exhibits two distinct bounce-back points. In stark contrast to the standard R-N model, these points are not situated behind the Cauchy horizon; rather, they are located entirely exterior to it. Of course, at this critical threshold, these two points are nearly degenerate, lying in close proximity to one another, as illustrated in (Fig.~\ref{4}).
\begin{figure}[htbp]
    \centering
    \includegraphics[width=1.01\linewidth, height=7.8cm]{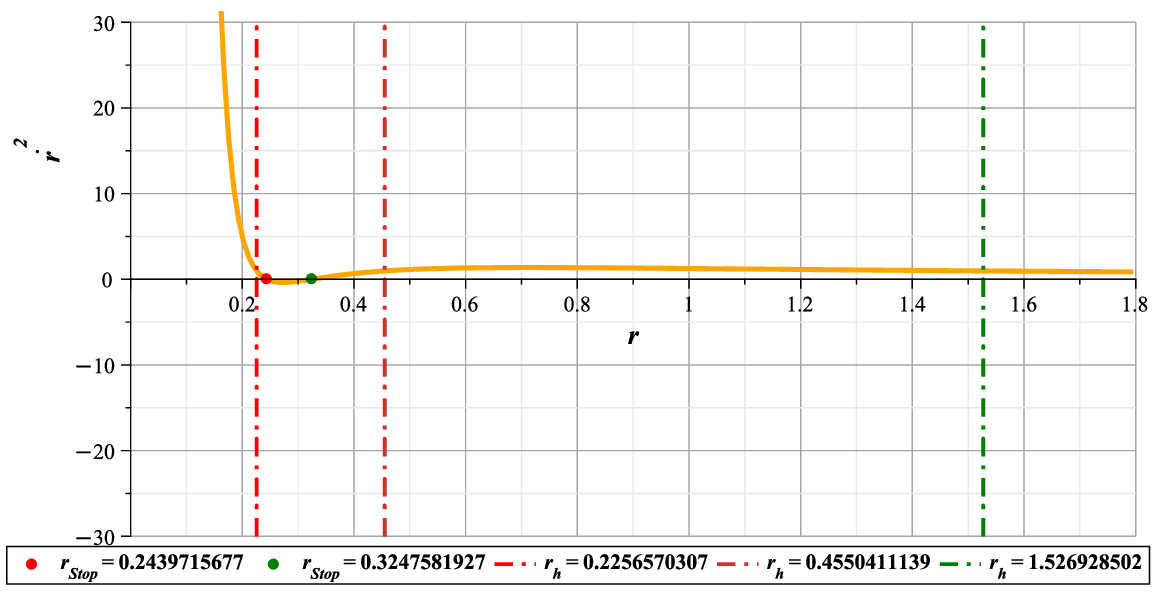}
     \hfill
     \includegraphics[width=1.01\linewidth, height=7.8cm]{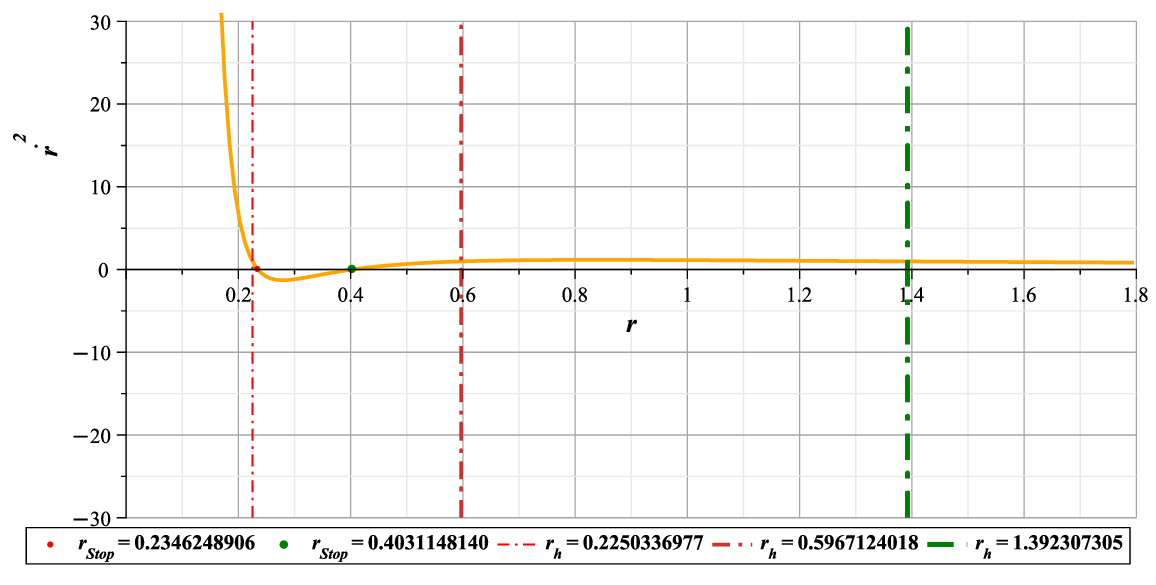}
    \hfill
    \includegraphics[width=1.01\linewidth, height=7.8cm]{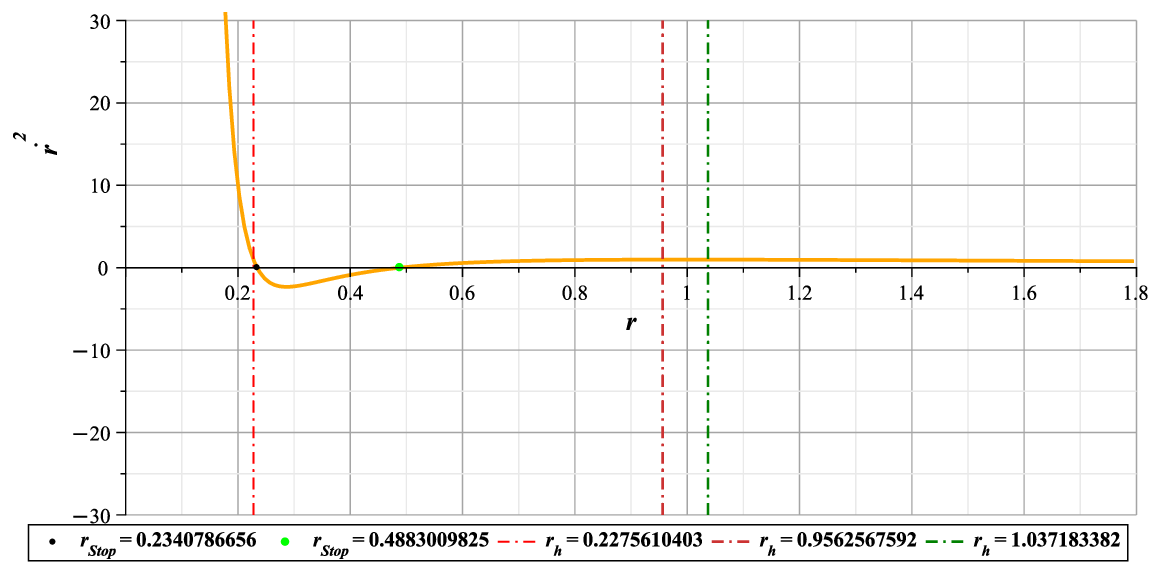}
    \hfill
        \caption{ $r^\cdot{^2}$  function with $\alpha_{2} = -0.004, b = 100, M = 1$ for 5a: $q=0.85$, 5b: $q=0.92$ , 5c: $q=1$  in III-Horizon structure of  NED Black hole.}
    \label{5}
\end{figure}

As the electric charge increases and the system approaches the extremal configuration, the intermediate horizon moves closer to the event horizon, while the separation between the two bounce-back points ($R_{\mathrm{stop}}$) increases (Fig.~\ref{5}). Further insight can be obtained by examining the radial equation of motion, Eq.~(\ref{eq:radial_velocity}). According to Eq.~(\ref{eq:radial_velocity}), only regions satisfying $\dot{r}^{\,2}\geq0$ are kinematically accessible to the particle, whereas regions with $\dot{r}^{\,2}<0$ are classically forbidden.)\\
As illustrated in Figs.~\ref{4} and \ref{5}, the appearance of two distinct $R_{\mathrm{stop}}$ points gives rise to a classically forbidden region ($\dot{r}^{\,2}<0$) between them, through which particle motion is not allowed. Both the width and the depth of this forbidden region increase with the black hole charge. Consequently, within this specific range of charge,, the nonlinear electromagnetic field generates an 'effective dynamical barrier' that prevents radially infalling particles from reaching the central singularity. (From a phenomenological perspective, this behavior resembles that of regular black hole models, such as the Bardeen and Hayward solutions. The underlying physical mechanism, however, is fundamentally different: in regular black holes the spacetime singularity is removed, whereas in the present model the singularity remains but is dynamically inaccessible to radially infalling particles within the corresponding parameter range.\\
Another notable feature, visible in Fig.~\ref{4} and more clearly in Fig.~\ref{5}, is that the bounce-back points are located between the intermediate and inner horizons. This location is significant for the following reasons:\\
\begin{enumerate}
\item \textbf{(i)}\\ In this region, which lies outside the innermost Cauchy-like horizon, the radial coordinate $r$ is spacelike. Consequently, the bounce-back corresponds to a genuine classical turning point within the same spacetime manifold, without invoking the maximal analytic extension to another asymptotically flat region.\\
\item \textbf{(ii)}\\  Although the innermost horizon may be susceptible to mass-inflation instabilities analogous to those associated with the Cauchy horizon of the RN spacetime, the intermediate horizon may possess a different causal character. Therefore, provided that the intermediate horizon is classically stable, the particle reaches the turning point and returns without crossing the innermost Cauchy-like horizon. Under this assumption, the bounce-back process can be interpreted within the same classical spacetime, without encountering the region expected to be affected by mass inflation.
\end{enumerate}
\subsubsection{ Toward Super-Extreme Charges}
As previously discussed, unlike the R-N black hole , which loses its event horizon and becomes a naked singularity when $q/M>1$, the present NED model preserves a distinct event horizon, thereby maintaining its black hole structure even in the presence of super-extremal charges. This unique feature motivated us to investigate the behavior of bounce-back points ($R_{\text{stop}}$) for neutral radially infalling particles in this regime.
\begin{figure}[htbp]
    \centering
    \includegraphics[width=1.01\linewidth, height=4.8cm]{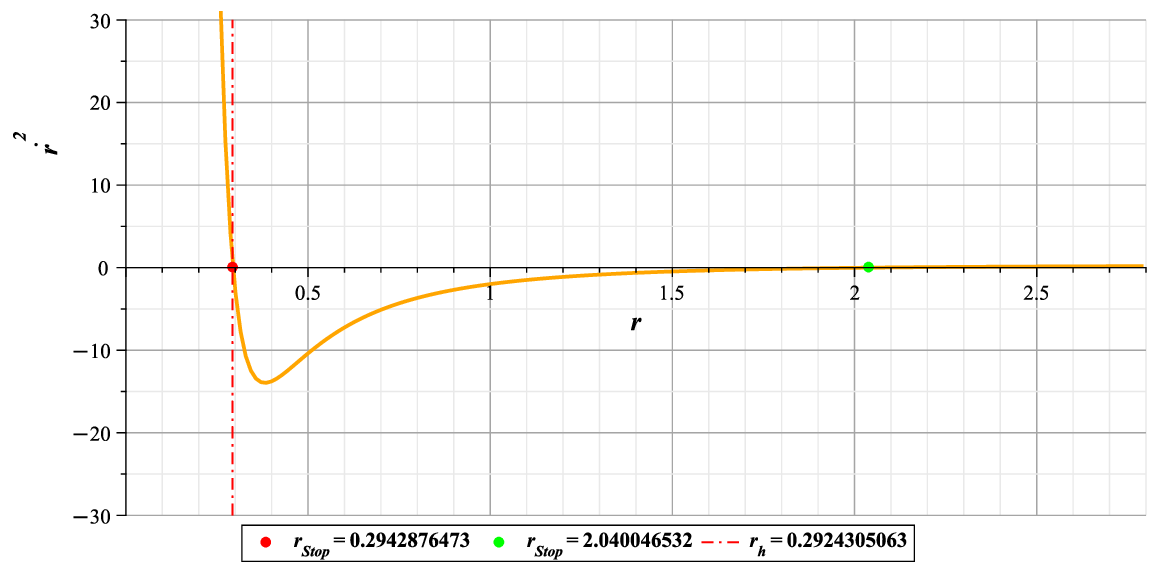}
     \hfill
     \includegraphics[width=1.01\linewidth, height=4.8cm]{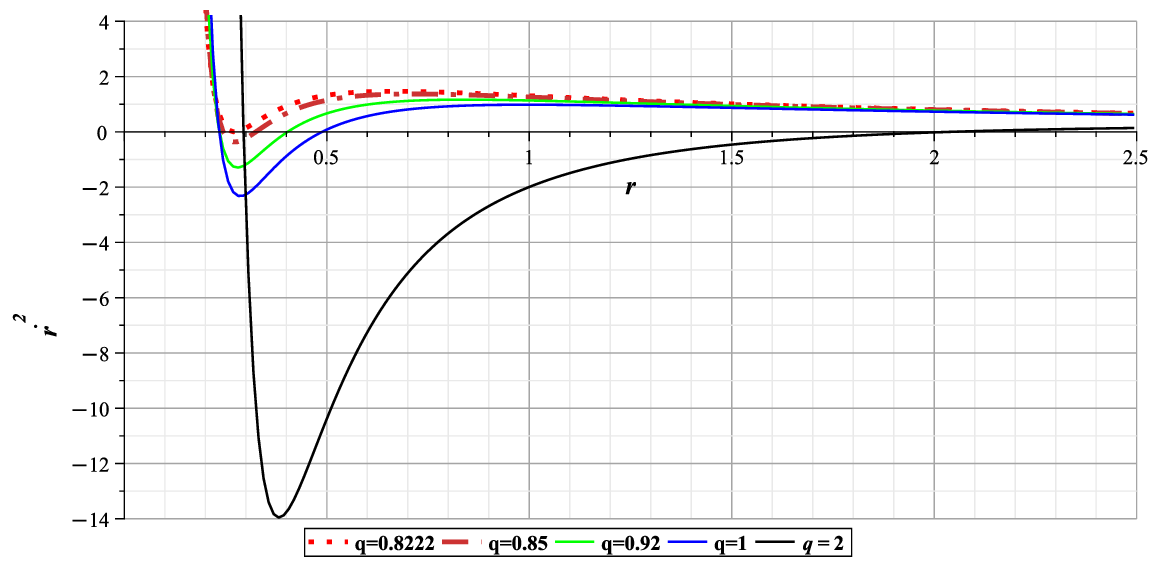}
    \hfill
\caption{ $r^\cdot{^2}$  function with $\alpha_{2} = -0.004, b = 100, M = 1$ for 6a: $q=0.85$, 6b: set of behaviors for different charges in III-Horizon structure of  NED Black hole.}
    \label{6}
\end{figure}
Our numerical calculations for various charge values reveal a fascinating nonlinear phenomenon. Contrary to monotonic behavior, the protective effect of the NED field depends strongly on the charge-to-mass ratio. Our results indicate that,  within a specific 'charge window' (for instance, $1 < q/M \lesssim 10$ for a particle released from $b=100M$), the dynamical forbidden region that forms in inner layers in the sub-extremal regime expands outward and eventually extends beyond the event horizon (Fig.~\ref{6}). In this range, the outer stopping point is located at a considerable distance from the horizon (e.g., $r \approx 2.04M$ for $q=2M$). Since this turning point lies outside the event horizon ($r>r_h$), particles released from rest at sufficiently large distances reverse their motion before crossing the horizon. Within this parameter range, the nonlinear electromagnetic field therefore acts as an effective dynamical barrier for neutral particles released from rest at large distances, suppressing their absorption by the black hole.\\
However, a remarkble point is that with further increase in the charge-to-mass ratio (e.g., $q/M \gtrsim 10$ for $b=100M$), the outer turning point continues to move outward until it eventually coincides with the particle's release position ($r=b$). Beyond this point, no additional turning point exists between the release position and the event horizon, allowing the particle to move toward the black hole. These results indicate that, for the present model and for particles released from rest at a fixed radius, there exists an 'optimal charge window' in which the absorption of neutral particles is suppressed by the nonlinear electromagnetic field. This behavior has no analogue in the Schwarzschild spacetime or in the RN solution in its extremality bound. The existence and extent of such a charge window are model dependent and should therefore be regarded as a important property of this NED solution.
 \subsection{ Tidal Force Behavioral Regime}
As discussed previously,  a positive tidal force indicates a tensile (repulsive) , whereas a negative value indicates a compressive (attractive) regime. The resulting competition between stretching and compression determines the tidal deformation experienced by an infalling body. In this section, we analyze the radial and angular components of the tidal force for the NED black hole using Eqs.~(\ref{eq:rad_force}) and (\ref{eq:ang_force}).
\subsubsection{ Radial Tidal force Behavior in NED Black Hole }
\begin{figure}[htbp]
    \centering
    \includegraphics[width=1.01\linewidth]{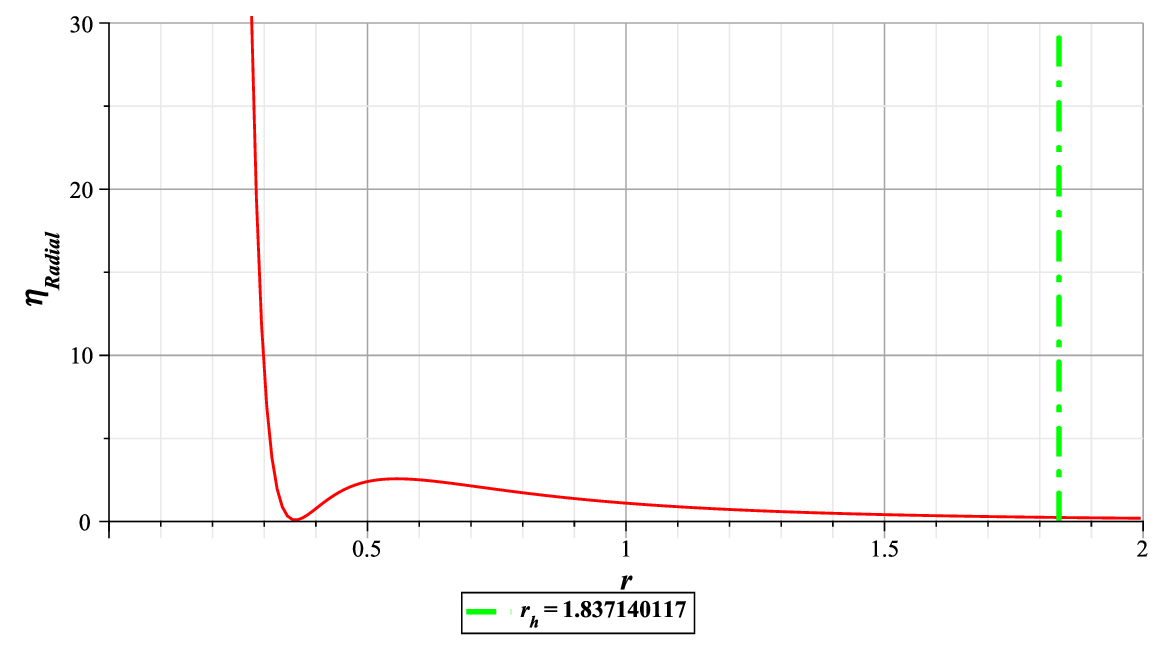}
    \caption{ $\eta_{Radial}$  function with $\alpha_{2} = -0.004, b = 100, M = 1, q = 0.547$ for III-Horizon structure of  NED Black hole.}
    \label{7}
\end{figure}

\begin{figure}[htbp]
    \centering
    \includegraphics[width=1.01\linewidth,height=7.8cm]{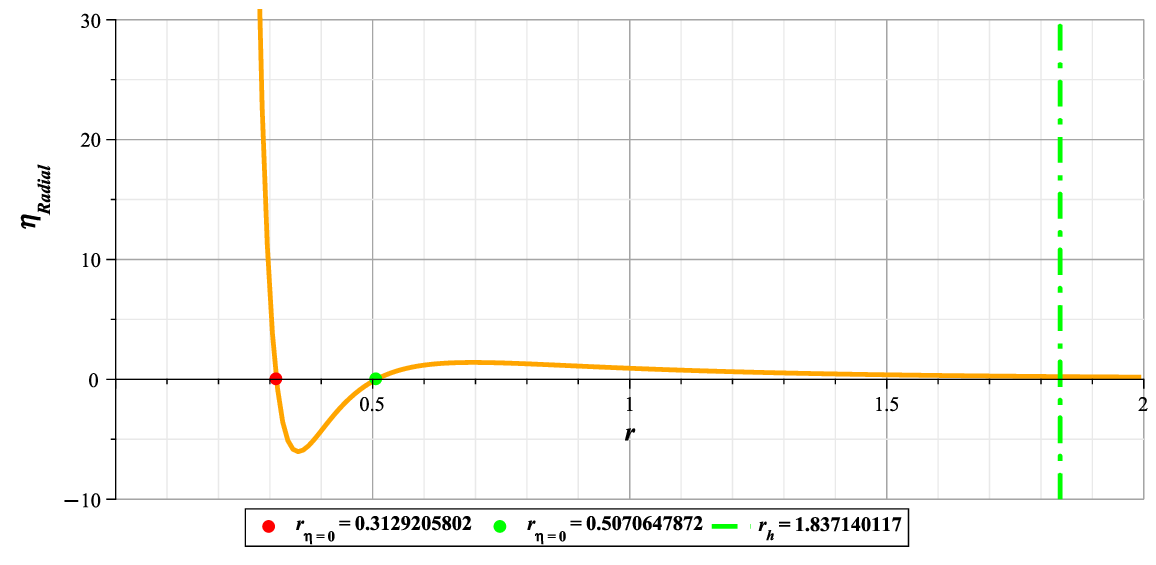}
     \hfill
    \includegraphics[width=1.01\linewidth,height=7.8cm]{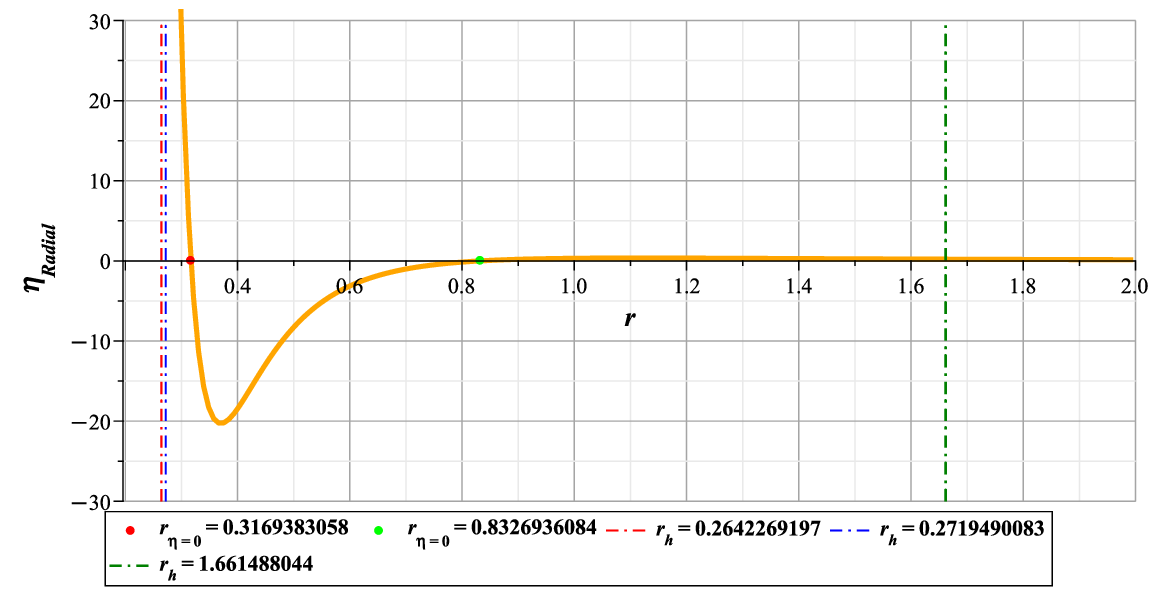}
     \hfill
     \includegraphics[width=1.01\linewidth,height=7.8cm]{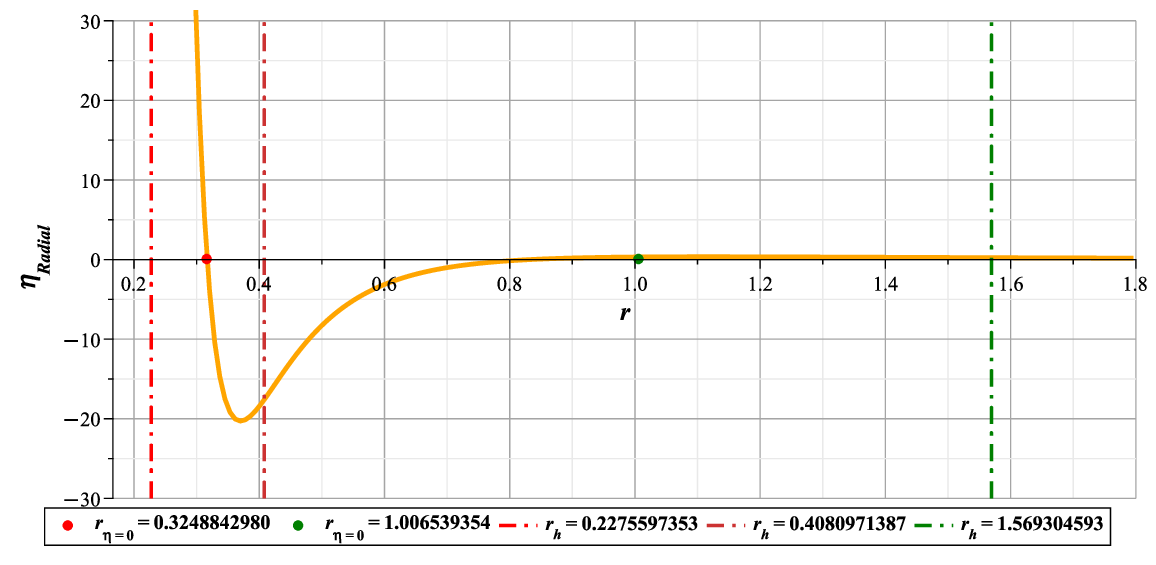}
    \hfill
\caption{ $\eta_{Radial}$  function with $\alpha_{2} = -0.004, b = 100, M = 1$ for 8a: $q=0.6$, 8b: $q=0.75$, 8c: $q=0.8222$  in III-Horizon structure of  NED Black hole.}
    \label{8}
\end{figure}

\begin{figure}[htbp]
    \centering
    \includegraphics[width=1.01\linewidth, height=5.8cm ]{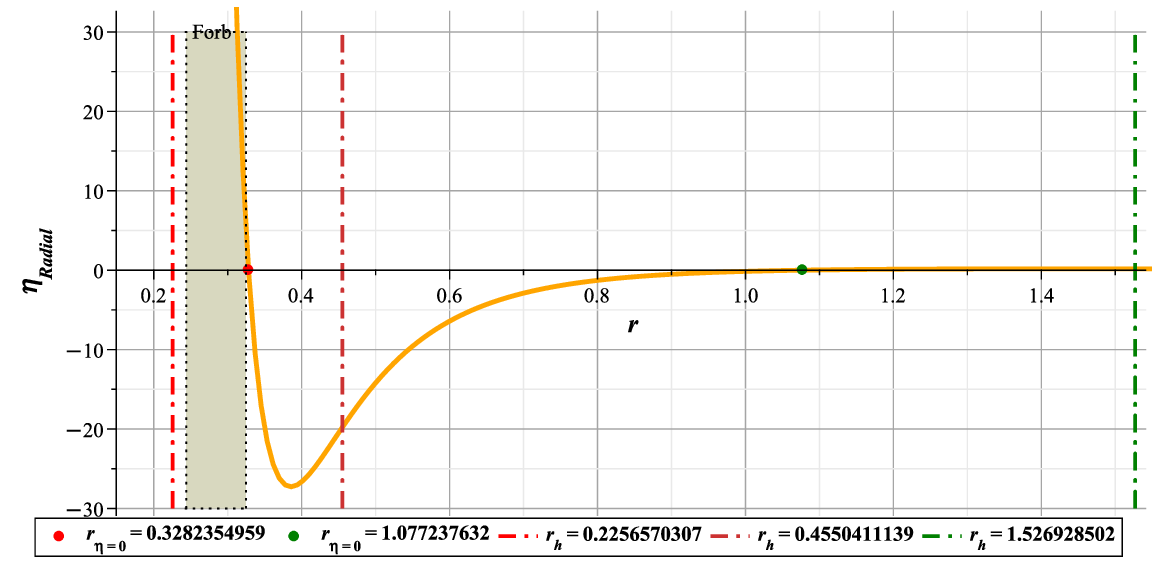}
    \hfill
         \caption{ $\eta_{Radial}$  function with $\alpha_{2} = -0.004, b = 100, M = 1, q=0.85$, in III-Horizon structure of  NED Black hole.}
    \label{9}
\end{figure}

Previous studies have shown that, in the Schwarzschild spacetime, the radial tidal force is always positive (stretching) and diverges as the singularity is approached. This infinite stretching exceeds the elastic limits of any physical structure, inevitably leading to the destruction of the infalling body (complete tidal disruption$\sim$ spaghettification). In the R-N spacetime, the presence of the electromagnetic field qualitatively modifies the radial tidal force, which changes sign at a specific radius and transitions from stretching to compression. Nevertheless, the compressive tidal force also diverges as the singularity is approached. Consequently, despite the change in sign, the ultimate fate of the infalling body remains tidal destruction ~\cite{9}.\\
For the NED black hole, our analysis shows that the radial tidal force exhibits three qualitatively distinct regimes in the sub-extremal parameter space, depending on the electric charge:\\ 
 \textbf{ Schwarzschild-Like Region ($q \leq 0.547$):}\\ Within this charge range, the radial tidal force remains positive throughout the spacetime and does not change sign. Although this behavior resembles that of the Schwarzschild solution, the nonlinear electromagnetic corrections generate nonzero local extrema in the tidal-force profile, distinguishing the present model from the Schwarzschild case (Fig.~\ref{7}).\\
 \textbf{New Behavioral Region ($0.547 < q < 0.8223$):} \\
Beyond the first threshold, the radial tidal force develops two zero crossings.For charge values corresponding to the single-horizon configuration, both zero crossings are located inside the event horizon. After the intermediate and inner horizons appear, the two zero crossings lie initially  between the event horizon and the intermediate horizon. As the charge is further increased, one of these points gradually moves into the region bounded by the intermediate and Cauchy horizons, as illustrated in Fig.~\ref{8}. Of course, this is a relative movement and it can be considered that with the movement of the intermediate horizon towards the event horizon, and also taking into account the net displacement of the second point itself, this point is placed between the intermediate horizon and the inner horizon.\\ 
Within this parameter range, where no nontrivial bounce-back point exists, the tidal-force profile differs qualitatively from both the Schwarzschild and R-N cases. An infalling body first experiences radial stretching after crossing the event horizon, followed by a transition to radial compression at the first zero crossing. At the second zero crossing, located close to the inner horizon, the radial tidal force changes sign once more, returning to a stretching regime and eventually diverging as the singularity is approached (Fig.~\ref{8}). Consequently, the infalling body is still expected to undergo complete tidal disruption in this regime.\\ 
\textbf{New R-N-Like Protected  Region} ($q > 0.8223$):\\
For $q>0.8223$, the radial dynamics enter a qualitatively different regime. The appearance of the 'dynamical forbidden region' modifies the trajectory of the infalling particle. For example, for $q=0.85$ (Fig.~\ref{9}), the particle experiences radial stretching after crossing the event horizon until approximately $r\simeq1.07$, after which the tidal force becomes compressive and remains so until the boundary of the classically forbidden region at approximately $r\simeq0.32$.
The key observation is that, according to the analysis presented in the previous subsection, the interval $0.24<r<0.32$ is a classically forbidden region, where $\dot{r}^{\,2}<0$. Since the inner boundary of this forbidden region lies close to the second zero crossing of the radial tidal force, the particle never reaches the region in which the tidal force returns to the stretching regime and subsequently diverges. Consequently, unlike the Schwarzschild and RN spacetimes, in which the radial tidal force diverges along the physically accessible trajectory, the particle in the present NED solution never encounters a divergent tidal force.

From a physical perspective, these results indicate that, within this parameter range, the nonlinear electromagnetic field effectively generates a region,\textit{protective shield}, that protects the infalling body from the divergent tidal forces encountered in the Schwarzschild and RN spacetimes. In other words, for particles whose motion is governed by the present solution, the nonlinear electromagnetic field can therefore prevent spaghettification by causing the particle to reverse its motion before it reaches the region of divergent tidal forces. 
\subsubsection{ Toward Super-Extreme Charges}
\begin{figure}[htbp]
\centering
\includegraphics[width=1.01\linewidth, height=5.2cm ]{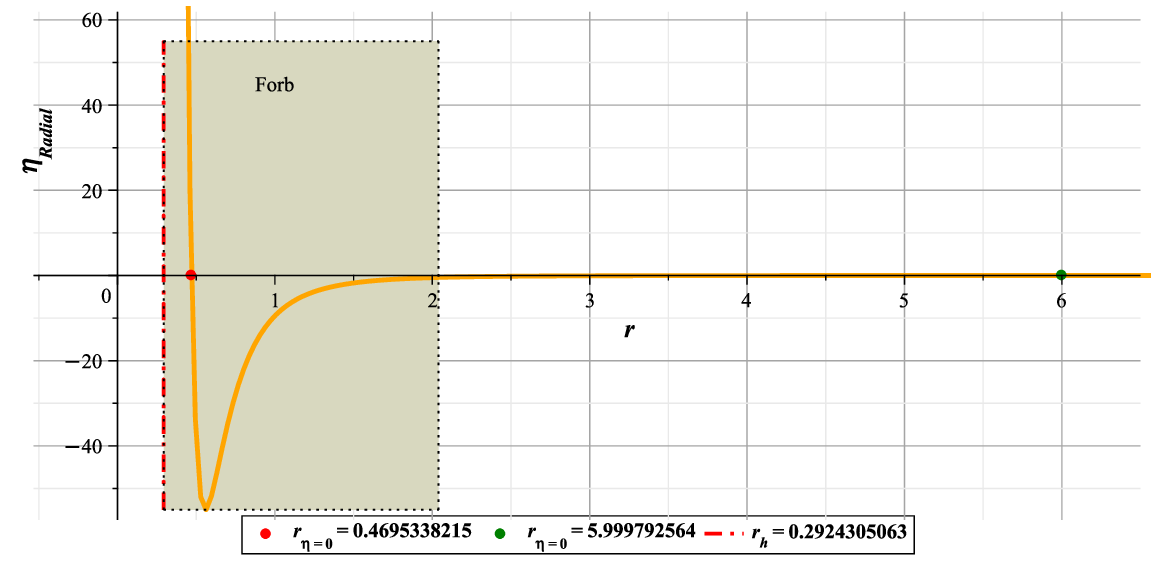}
\hfill
\includegraphics[width=1.01\linewidth,height=5.2cm ]{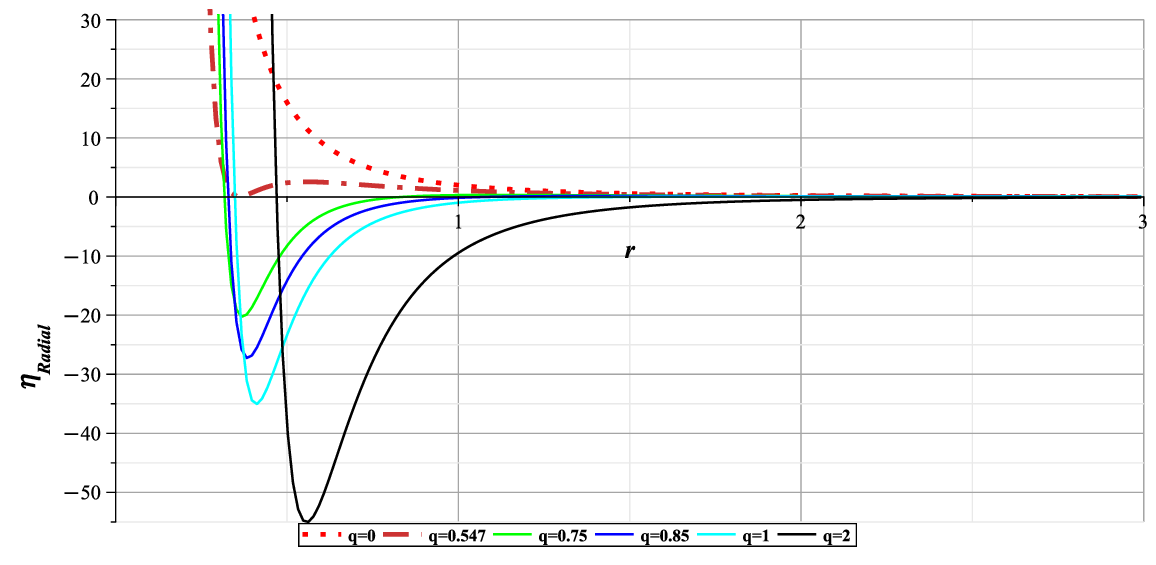}
\hfill
\caption{ $\eta_{Radial}$  function with $\alpha_{2} = -0.004, b = 100, M = 1$ for 10a: $q=2$, 10b: set of behaviors for different charges in III-Horizon structure of  NED Black hole.}
\label{10}
\end{figure}

The behavior of the radial tidal force in the super-extremal regime combines the features discussed previously for the bounce-back points and the protective behavior associated with the classically forbidden region. In this regime, two distinct protective mechanisms operate simultaneously. As shown in Fig.~\ref{10}, the radial tidal force has two zero crossings: an inner one at $r_{\eta=0}^{\mathrm{inner}}\approx0.47$, close to the event horizon ($r_h\approx0.29$), and an outer one at $r_{\eta=0}^{\mathrm{outer}}\approx6.0$.\\ The region between these zero crossings, together with the classically forbidden region (shaded area), plays a central role in the particle dynamics. As shown previously, neutral particles released from rest at sufficiently large distances cannot penetrate the classically forbidden region. Instead, they reach its outer boundary (approximately $r\simeq2.0$ for a particle released from $b=100$), where they reverse their motion and move outward.\\ \textit{Consequently, such particles neither cross the event horizon nor encounter the strong tidal fields associated with the central singularity.}
A second protective mechanism, is even more remarkable, arises from the behavior of the radial tidal force itself. The radial tidal force changes sign at $r \approx 6.0M$, transitioning from stretching (positive) to compressing (negative). Unlike the RN spacetime discussed in Ref.~\cite{9}, in which the tidal force diverges along the physically accessible trajectory, the radial tidal force in the present model remains finite throughout the region accessible to the particle.\\ This dual protection mechanism, combining the forbidden region barrier with bounded tidal forces, represents a significant remarkable feature in NED model in the super-extremal regime that can fundamentally alter the fate of infalling our particle in extreme gravitational environment.
 \subsection{ Angular Tidal force Behavior in NED Black Hole } 
\begin{figure}[htbp]
    \centering
    \includegraphics[width=1.01\linewidth, height=5.8cm ]{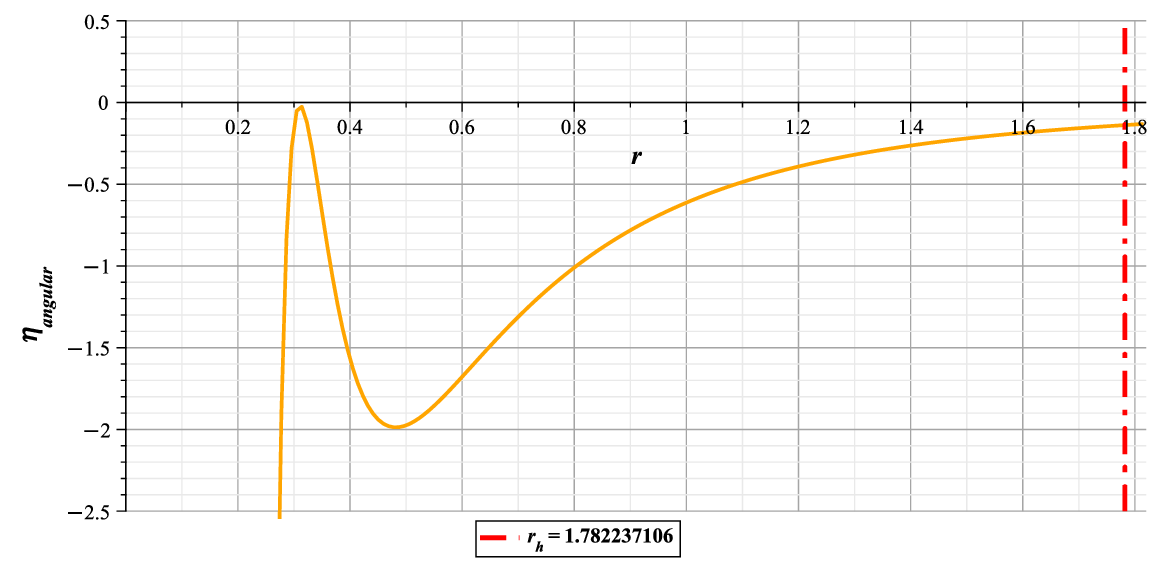}
    \hfill
         \caption{ $\eta_{angular}$  function with $\alpha_{2} = -0.004, b = 100, M = 1, q=0.623$, in III-Horizon structure of  NED Black hole.}
    \label{11}
\end{figure}

\begin{figure}[htbp]
    \centering
        \includegraphics[width=1.01\linewidth,height=7.3cm]{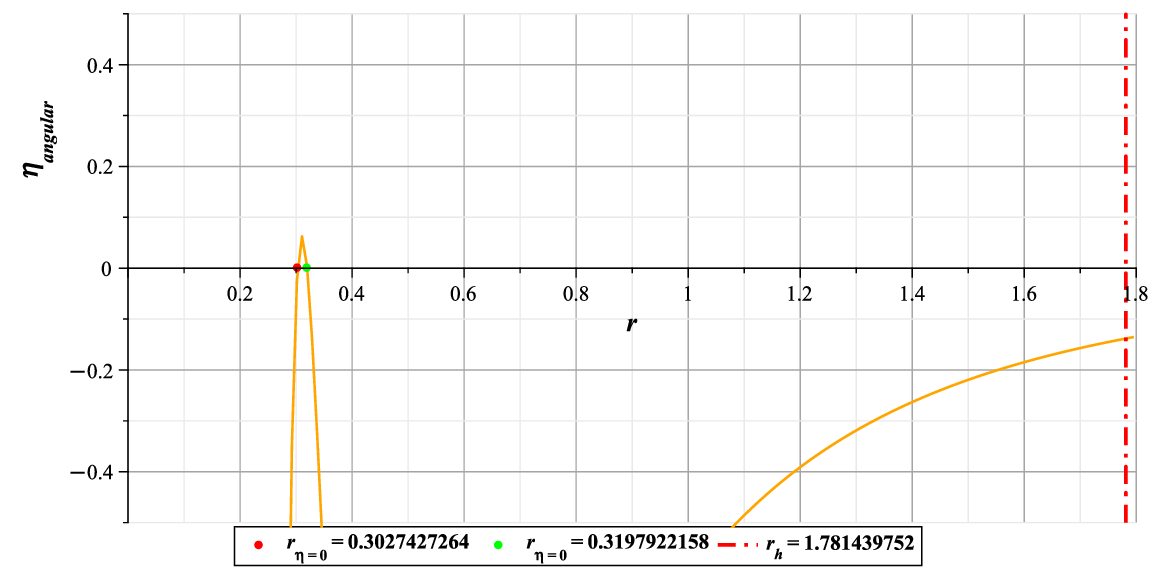}
     \hfill
     \includegraphics[width=1.01\linewidth,height=7.3cm]{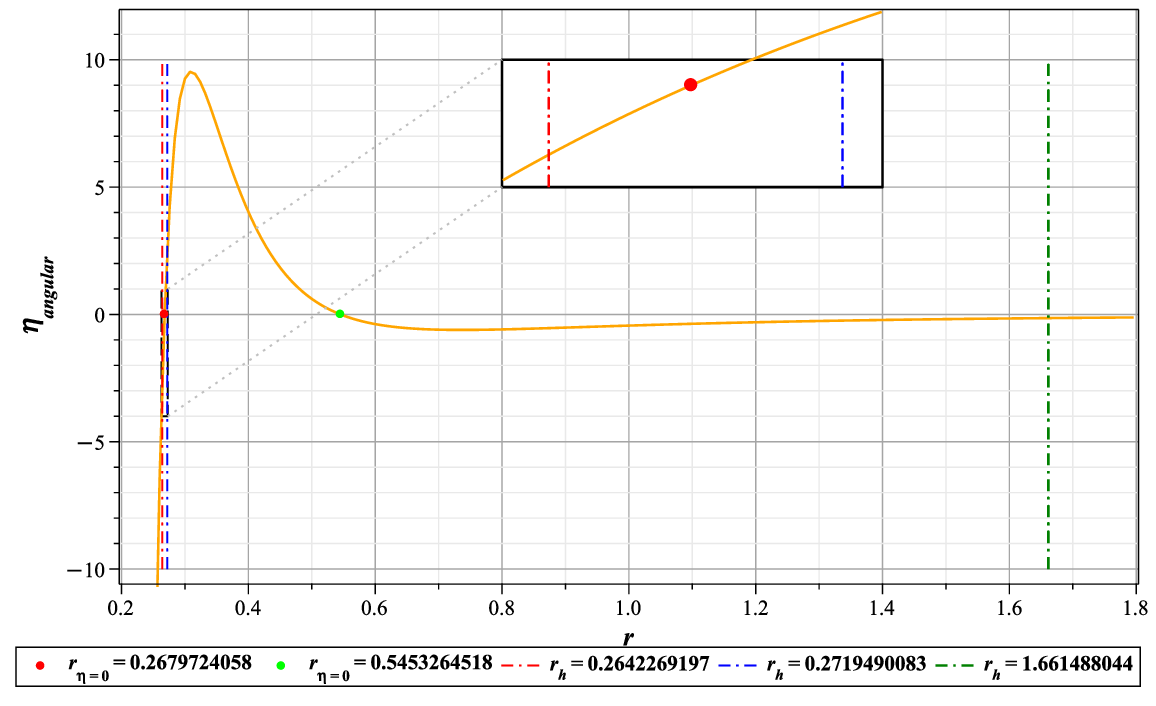}
    \hfill
\caption{ $\eta_{Radial}$  function with $\alpha_{2} = -0.004, b = 100, M = 1$ for 11a: $q=0.624$, 11b: $q=0.75$, in III-Horizon structure of  NED Black hole.}
    \label{12}
\end{figure}

\begin{figure}[htbp]
    \centering
    \includegraphics[width=1.01\linewidth, height=5.8cm ]{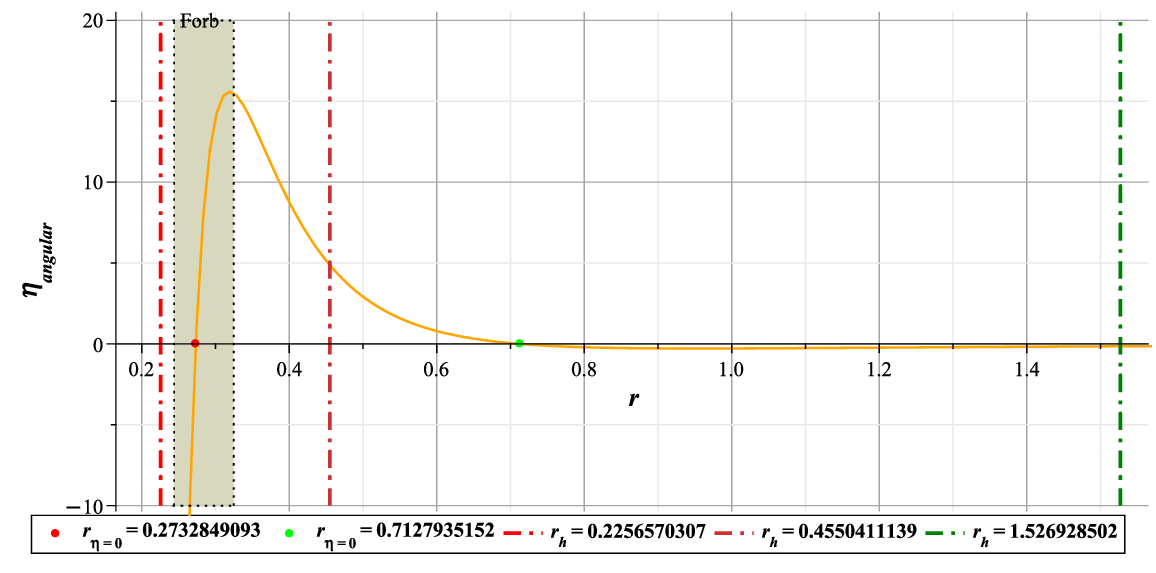}
    \hfill
    \includegraphics[width=1.01\linewidth, height=5.8cm ]{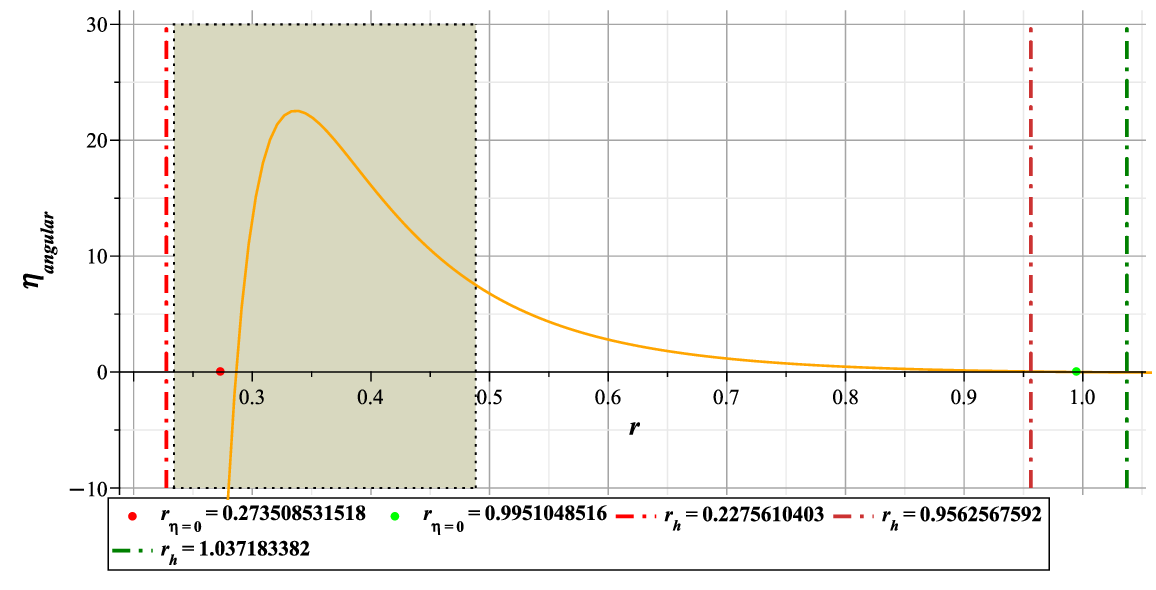}
    \hfill
         \caption{ $\eta_{angular}$  function with $\alpha_{2} = -0.004, b = 100, M = 1$ for 13a: $q=0.85$, 13b: $q=1$ in III-Horizon structure of  NED Black hole.}
    \label{13}
\end{figure}

In the Schwarzschild spacetime, the angular tidal force is always negative, corresponding to angular compression. Its magnitude also diverges as the singularity is approached, resulting in complete transverse compression of an infalling body. The introduction of the electromagnetic field in the R-N spacetime, also qualitatively modifies this behavior. Similar to the radial component, the angular tidal force in the R-N spacetime, also changes sign and transitioning from compression to stretching. The divergence at the end of the trajectory remains a common feature for both force components in the RN black hole ~\cite{9}.\\
Our analysis of the angular tidal force over a broad range of charge values reveals a behavior that differs qualitatively from both the Schwarzschild and RN spacetimes while exhibiting several similarities to the radial tidal force in the present NED solution. As for the radial component, three distinct regimes can be identified according to the value of the black hole charge $q$. The corresponding threshold values, however, differ from those found for the radial tidal force.  
  \textbf{ Schwarzschild-Like Region ($q \leq 0.623$):}\\ 
 Our calculations indicate that within this range, the angular component of the tidal force remains negative throughout the entire spacetime and exhibits no sign change. However, the presence of non-zero extrema (maxima and minima) arising from the NED term distinguishes our model from a pure Schwarzschild black hole (Fig.~\ref{11}).\\ 
\textbf{New Behavioral Region ($0.623 < q < 0.8223$):} \\  
For the angular tidal force, two zero crossings first appear inside the event horizon while the spacetime still possesses a single horizon. After the intermediate and inner horizons emerge, the qualitative evolution closely follows that observed for the radial tidal force, as illustrated in Fig.~\ref{12}.\\
Within this regime, an infalling body first experiences angular compression after crossing the event horizon. At the first zero crossing, the angular tidal force changes sign and becomes stretching. At the second zero crossing, located close to the inner horizon, the angular tidal force changes sign once again, returning to compression and eventually diverging as the singularity is approached (Fig.~\ref{12}). Consequently, in this regime, the infalling body still suffers structural disruption.\\  
\textbf{New  R-N-Like Protected  Region} ($q > 0.8223$):\\  
 As shown in the Fig.~\ref{13}, fo $q=0.85$ the angular tidal force exhibits two zero points: one at $r_{\eta=0}^{\text{outer}} = 0.71279$ (located between the event horizon at $r_h = 1.5269$ and the intermediate horizon at $r_h = 0.4550411139$) and another at $r_{\eta=0}^{\text{inner}} = 0.2732849093$ (situated between the intermediate and inner horizons). 
 A particle falling from the exterior first experiences a compressive angular force after crossing the event horizon. At $r \approx 0.713M$, the force changes sign and becomes stretching. The subsequent evolution, however, differs qualitatively from that of the Schwarzschild and R-N spacetimes. The region $0.244M < r < 0.325M$ constitutes a classical forbidden region ($\dot{r}^2 < 0$). Crucially, the inner zero point of the angular force ($r_{\eta=0}^{\text{inner}} \approx 0.273M$) lies \textit{within} this forbidden region. Since the particle cannot penetrate this region, it never reaches the second zero crossing and therefore never encounters the compressive tidal force that subsequently diverges as the singularity is approached. Instead, the particle reaches the outer boundary of the forbidden region, where its radial velocity vanishes, and then reverses its motion. Within the present NED solution, the nonlinear electromagnetic field therefore prevents the particle from entering the region in which destructive angular compression would occur in the Schwarzschild or RN spacetimes. A similar interpretation applies to the case $q=1$. The only qualitative difference is that the outer zero crossing is located outside the event horizon (Fig.~\ref{13}), a transition that is also observed in the RN spacetime.\\
 This mechanism complements the behavior of the radial tidal force. In both cases, the particle reverses its motion before reaching the region where the tidal forces would diverge, thereby avoiding the divergent tidal stresses present in the Schwarzschild and R-N spacetimes. For particles following the trajectories considered here, the nonlinear electromagnetic field therefore modifies the tidal evolution by preventing the onset of divergent radial and angular tidal forces responsible for spaghettification.
\subsubsection{ Toward Super-Extreme Charges}
\begin{figure}[htbp]
    \centering
    \includegraphics[width=1.01\linewidth, height=5.8cm ]{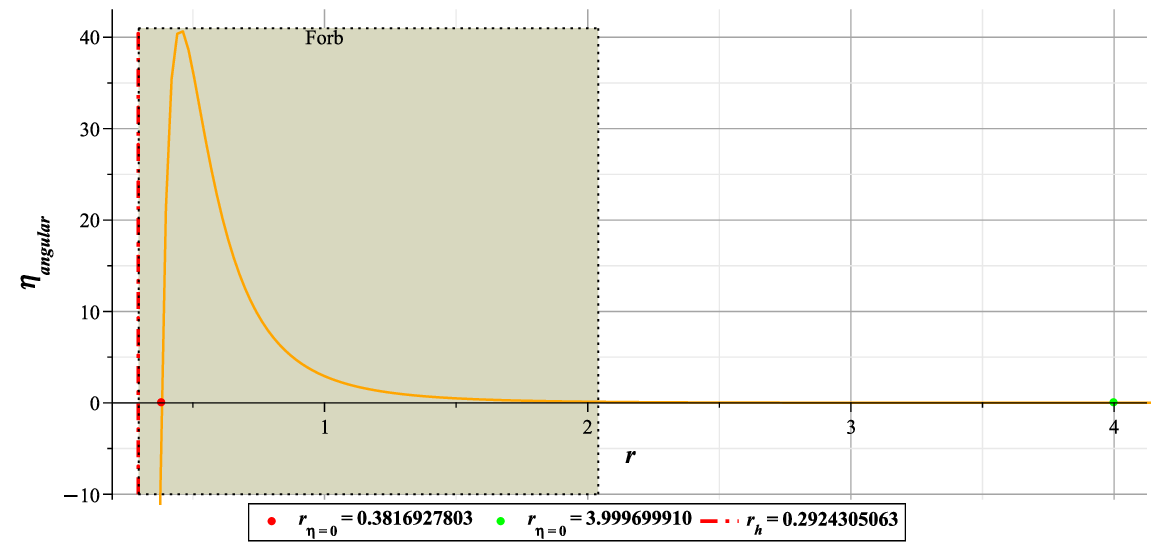}
    \hfill
    \includegraphics[width=1.01\linewidth, height=5.8cm ]{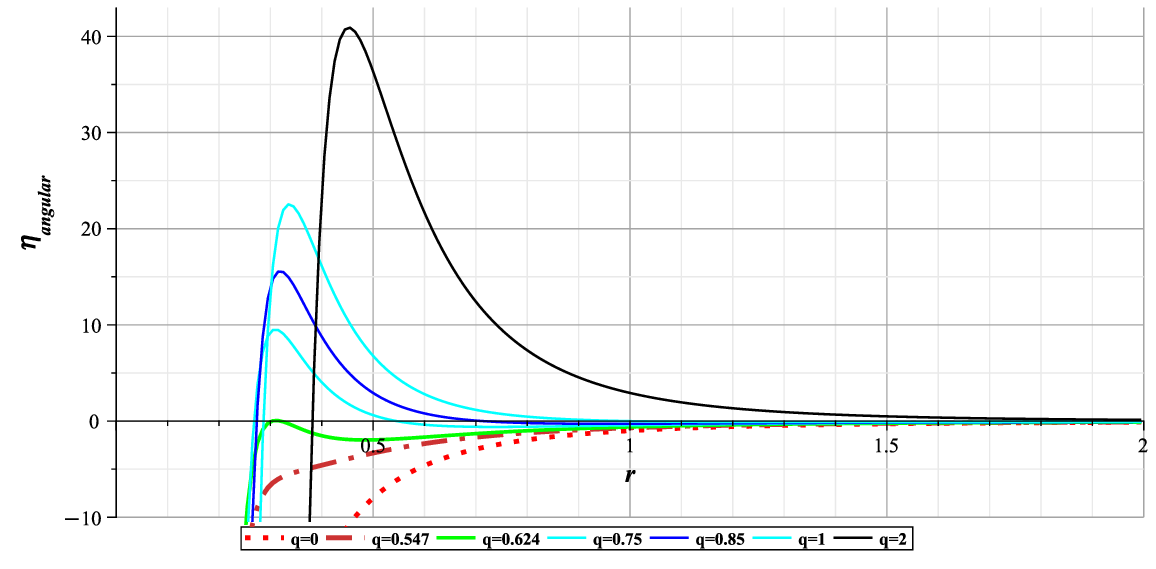}
    \hfill
         \caption{ $\eta_{angular}$  function with $\alpha_{2} = -0.004, b = 100, M = 1$ for 14a: $q=2 $ and 14b: set of behaviors for different charges in III-Horizon structure of  NED Black hole.}
    \label{14}
\end{figure}

In the super-extremal regime, the two protective mechanisms identified for the radial tidal force remain present. The angular tidal force again exhibits two zero points (Fig. ~\ref{14}). 
As discussed previously, neutral particles released from rest at sufficiently large distances cannot penetrate the shaded classically forbidden region. Instead, they reach its outer boundary, where they reverse their motion and move outward. Consequently, these particles neither cross the event horizon nor encounter the strong tidal fields associated with the central singularity. This dual protection mechanism, which combines the forbidden zone barrier with limited tidal forces, represents a remarkable and important feature for our NED model.
 \subsection{ Geodesic Deviation Equations} 
 Given the general form of the metric(\ref{eq:metric}) and energy function, substituting \ref{eq:rad_force} and \ref{eq:ang_force} into \ref{eq:deviation} will result in two differential equations:
 \begin{eqnarray}
  (\mathcal{E}^2-\mathfrak{f}(r))\xi^{\hat{1}\prime\prime}-\frac{\mathfrak{f}^{\prime}(r)}{2} \xi^{\hat{1}\prime} +
  \frac{\mathfrak{f}^{\prime\prime}(r)}{2} \xi^{\hat{1}}&=& 0,\\
  (\mathcal{E}^2-\mathfrak{f}(r))\xi^{\hat{i}\prime\prime}-\frac{\mathfrak{f}^{\prime}(r)}{2} \xi^{\hat{i}\prime} +
  \frac{\mathfrak{f}^{\prime\prime}(r)}{2} \xi^{\hat{i}}&=& 0. 
\end{eqnarray}  
 where $i = (2, 3)\equiv(\theta,\phi)$. The general solution to this set of equations for our case is as follows:
 \begin{equation*}
\kappa \! \left(r \right)=-\frac{2 M}{b}+\frac{q^{2}}{b^{2}}+\frac{2 \alpha_{2} q^{4}}{5 b^{6}}+\frac{2 M}{r}-\frac{q^{2}}{r^{2}}-\frac{2 \alpha_{2} q^{4}}{5 r^{6}},
\end{equation*}
\begin{equation}
\begin{split}
&\xi^{\hat{1}}(r) = \sqrt{\kappa(r)} \frac{5 b^{7} \dot{\xi}^{\hat{1}}(b)}{5 M b^{5} - 5 b^{4} q^{2} - 6 \alpha_{2} q^{4}} +\\ & \sqrt{\kappa(r)} \left( \frac{M}{b^{2}} - \frac{q^{2}}{b^{3}} - \frac{6 \alpha_{2} q^{4}}{5 b^{7}} \right) \xi^{\hat{1}}(b) \int \frac{dr}{\kappa(r)^{3/2}} \\,
\end{split}
 \end{equation} 
 \begin{equation}
\xi^{\hat{i}}=r \left(\frac{\xi^{\hat{i}}(b)}{b}-b \dot{\xi}^{\hat{i}}(b) \int \frac{1}{\sqrt{\kappa(r)}} dr \right).
\end{equation}  
 \section{ Tidal Forces and the IV-Horizon structure} 
 In this case, our metric function is given by ~\cite{67}: 
 \begin{equation}  
f \! \left(r \right)=1-\frac{2 M}{r}+\frac{q^{2}}{r^{2}}+\frac{2 q^{4} \alpha_{2}}{5 r^{6}}+\frac{q^{6} \left(16 \alpha_{2}^{2}-4 \alpha_{3}\right)}{9 r^{10}}  
\end{equation}  
We performed the same analysis carried out for the three-horizon solution over a broad range of black hole charge values for the four-horizon metric. To avoid unnecessary repetition, we will summarize the features that are very similar to the results of the three-horizon solution. Instead, we will focus primarily on describing and elaborating in detail any differences that may arise.  
\subsection{ $R_{\text{stop}}$ Emergence  Behavioral Regime}  
To facilitate a direct comparison with the three-horizon solution, we choose the model parameters so that the resulting configurations are as similar as possible. As before, we consider a neutral test particle released from rest at $r=b=100$, well outside the event horizon, and study its radial motion. Throughout this section we fix $\alpha_{2}=-0.0004$, $\alpha_{3}=6.38736\times10^{-7}$, and $M=1$.) As in the three-horizon solution, distinct charge thresholds separate the different regimes associated with the appearance of bounce-back points and changes in the tidal-force profile. 
The first point that should be noted in studying this configuration is that, for the chosen parameter values, the sub-extremal solution always possesses two horizons before the additional inner horizons appear. A notable difference is that, for $q<0.538$, before the four-horizon configuration is established, the radial dynamics resemble the RN solution more closely than the Schwarzschild solution, in contrast to the behavior observed for the three-horizon model.\\
\begin{figure}[htbp]
    \centering
    \includegraphics[width=1.01\linewidth, height=5.8cm]{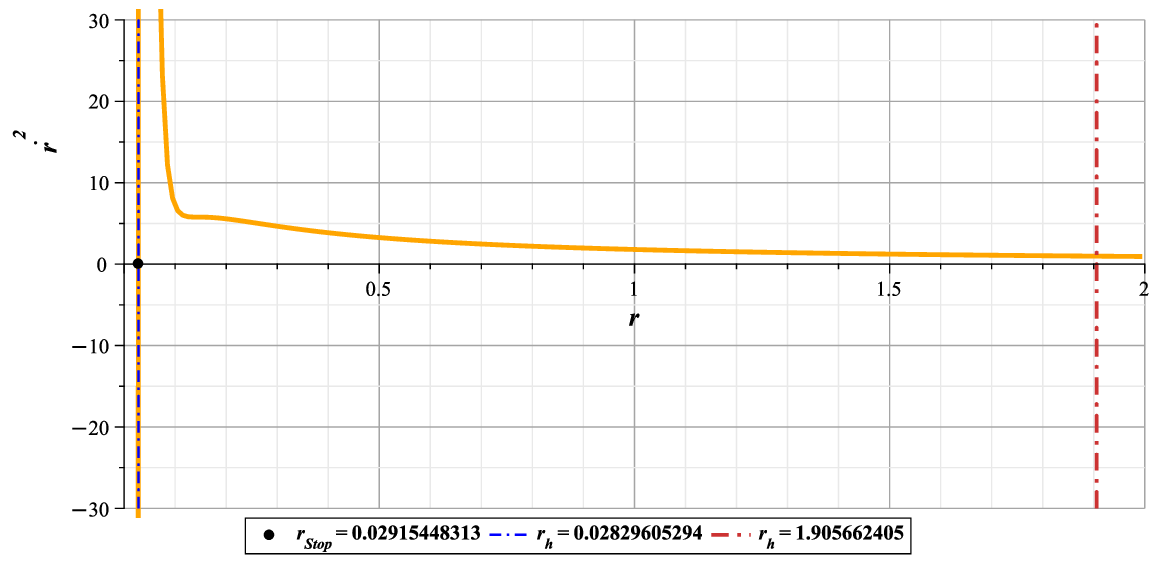}
     \hfill
     \includegraphics[width=1.01\linewidth, height=5.8cm]{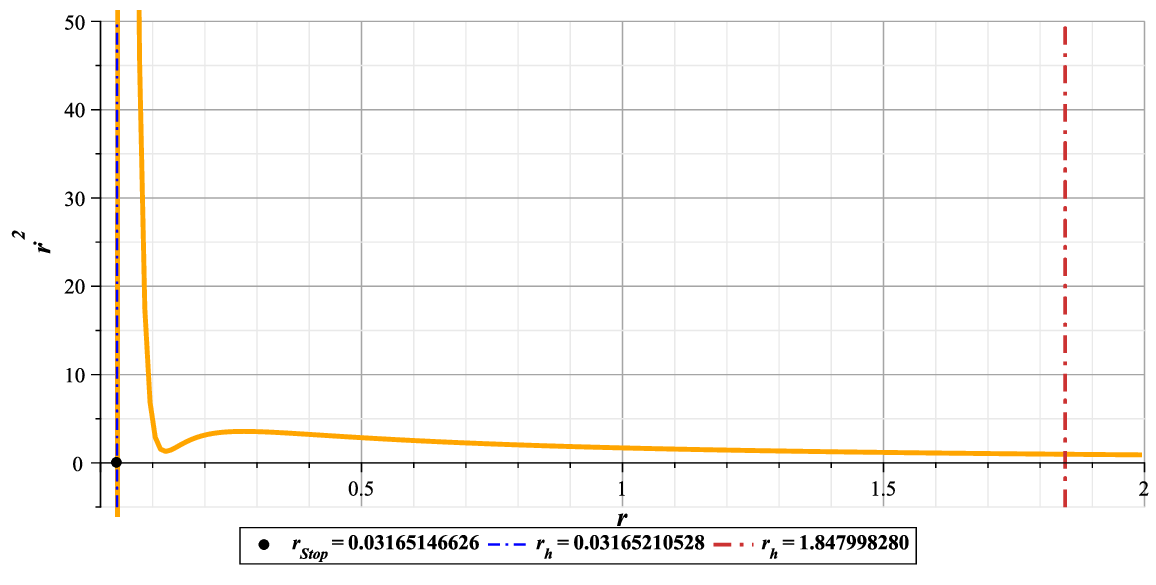}
    \hfill
\caption{ $r^\cdot{^2}$  function with $\alpha_{3} = 6.38736 \times 10^{-7},\alpha_{2} = -0.004, b = 100, M = 1$ for 6a: $q=0.424$, 6b: $q=0.53$ in VI-Horizon structure of  NED Black hole.}
    \label{15}
\end{figure} 
This similarity should, however, be interpreted with caution. Unlike the RN case, the bounce-back point remains located very close to the innermost horizon and exhibits only a very weak dependence on the black hole charge.  Furthermore, although this point is apparently accessible from a mathematical point of view, the large magnitude of the changes in the motion towards this point, as shown in Fig. ~\ref{15}, on the one hand and the properties of the Cauchy horizon and the problems associated with mass inflation in reaching or crossing this horizon on the other hand,  make this bounce-back point physically limited in its accessibility and interpretability.  
This single-point trend is maintained even with the emergence of different behavioral layers of space-time (the emergence of intermediate horizons) up to the boundary of $q=0.56$, for $q\geq0.561$, however, two additional bounce-back points appear, so that the radial motion is characterized by three turning points. However, as in the three-horizon solution, the appearance of these additional turning points is accompanied by the formation of a classically forbidden region, which renders the two innermost turning points inaccessible to particles following the trajectories considered here (Fig.~\ref{16}).
\begin{figure}[htbp]
    \centering
    \includegraphics[width=1.01\linewidth, height=5.8cm]{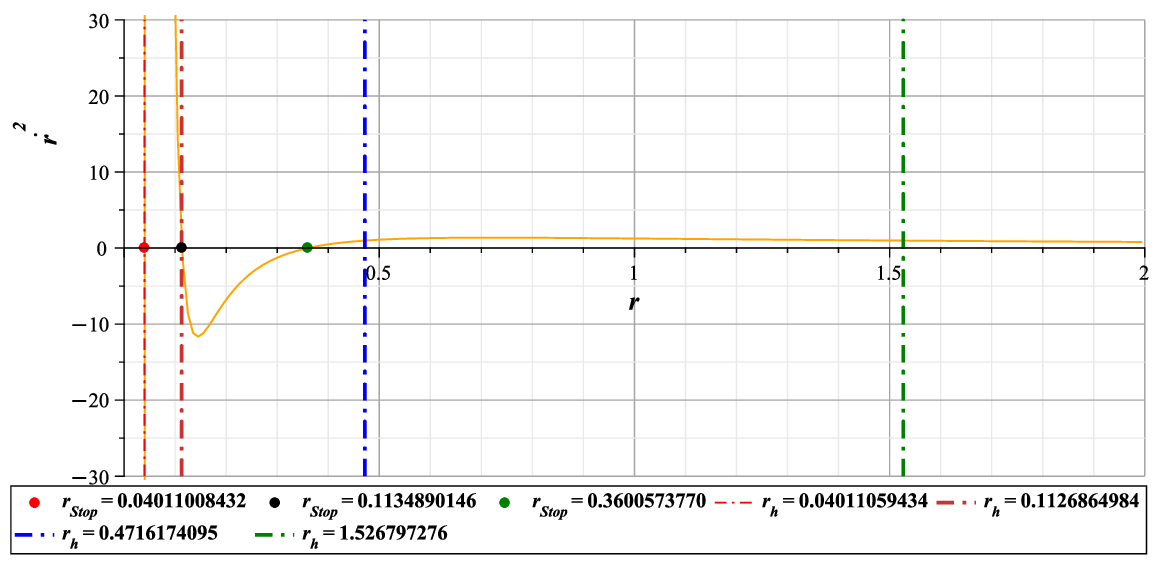}
     \hfill
     \caption{ $r^\cdot{^2}$  function with $\alpha_{3} = 6.38736 \times 10^{-7},\alpha_{2} = -0.004, b = 100, M = 1$, $q=0.85$ in VI-Horizon structure of  NED Black hole.}
    \label{16}
\end{figure}   
As in the three-horizon solution, the four-horizon geometry continues to admit black hole solutions beyond the R-N extremality bound. In this case as well, in the super-extremal regime, the classically forbidden region expands outward and extends beyond the event horizon. Consequently, particles released from rest at sufficiently large distances reverse their motion before crossing the event horizon. Within this parameter range, the nonlinear electromagnetic field therefore acts as an effective dynamical barrier for the class of trajectories considered here (Fig.~\ref{17}).  
\begin{figure}[htbp]
    \centering
    \includegraphics[width=1.01\linewidth, height=5.8cm]{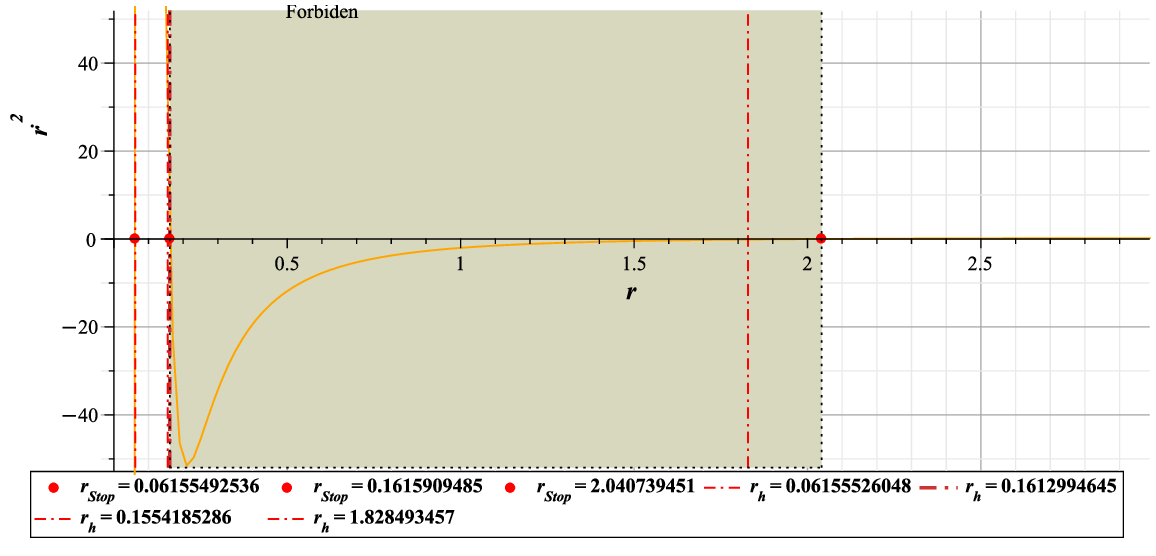}
     \hfill
     \caption{ $r^\cdot{^2}$  function with $\alpha_{3} = 6.38736 \times 10^{-7},\alpha_{2} = -0.004, b = 100, M = 1$, $q=2$ in VI-Horizon structure of  NED Black hole.}
    \label{17}
\end{figure}    
 \subsection{ Tidal Force Behavioral Regime}  
The radial and angular components of the sub extremal NED black hole in the four-horizon form exhibit a similar behavior to the three-horizon form with a slight deviation, meaning that three behavioral regions can be identified in both components, although with different ranges of black hole charge compared to the three-horizon case.\\  
 \textbf{Radial Tidal force}
 \begin{itemize}
    \item Schwarzschild-Like Region: $ q\leq 0.373$.
    \item New Behavioral Region: $0.373 < q < 0.561$.
    \item R-N Like Protected Region: $ q\geq 0.561$.
\end{itemize} 
 \textbf{Angular Tidal force}
 \begin{itemize}
    \item Schwarzschild-Like Region:$ q\leq 0.4244 $.
    \item New Behavioral Region: $0.4244 < q < 0.561$.
    \item  R-N Like Protected Region: $ q\geq 0.561$.
\end{itemize} 
It is worth emphasizing that the phrase 'with slight deviation' has a different interpretation in each of the three behavioral regimes. In all these configurations, there consistently exists a zero point of the force component,which leads to a sign change and consequently alters the force behavior, located near the innermost horizon, and this point does not exhibit significant displacement as the charge parameter is varied.
 \begin{figure}[htbp]
    \centering
    \includegraphics[width=1.01\linewidth,height=7.3cm]{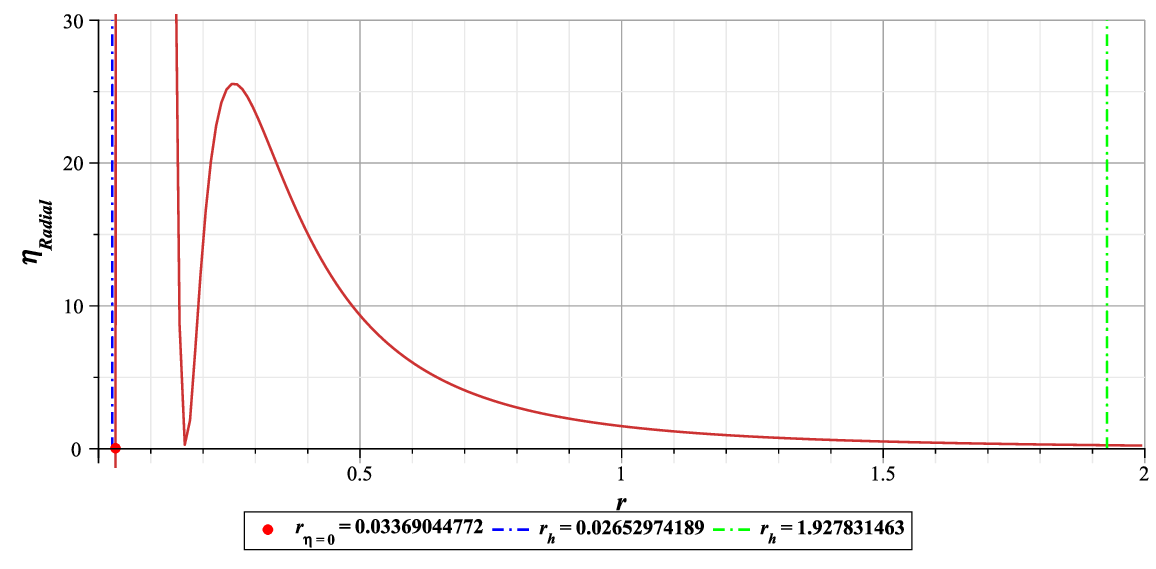}
     \hfill
    \includegraphics[width=1.01\linewidth,height=7.3cm]{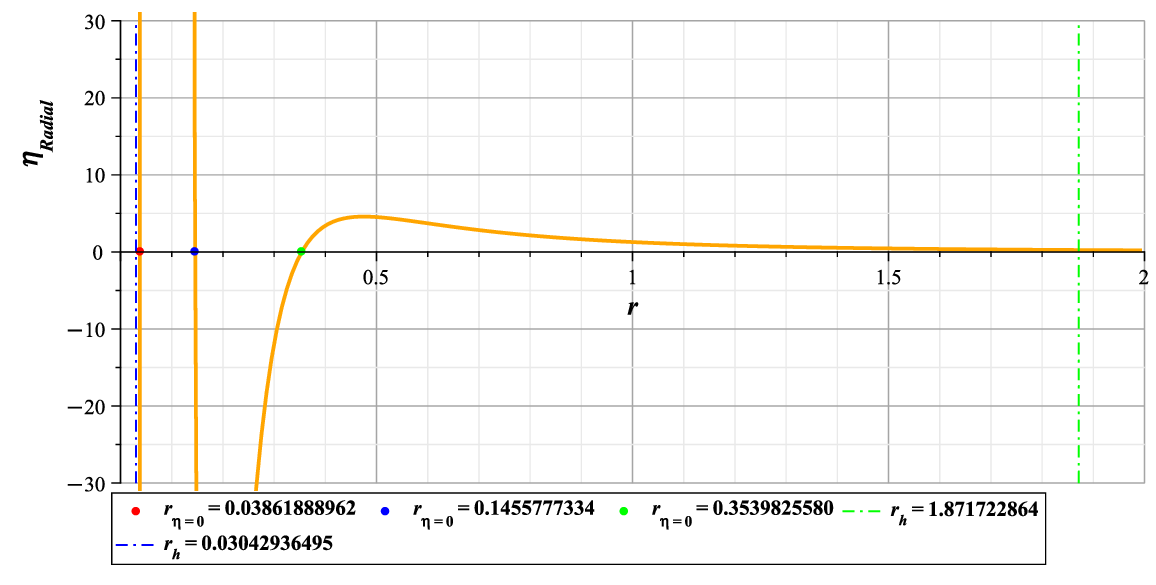}
     \hfill
     \includegraphics[width=1.01\linewidth,height=7.3cm]{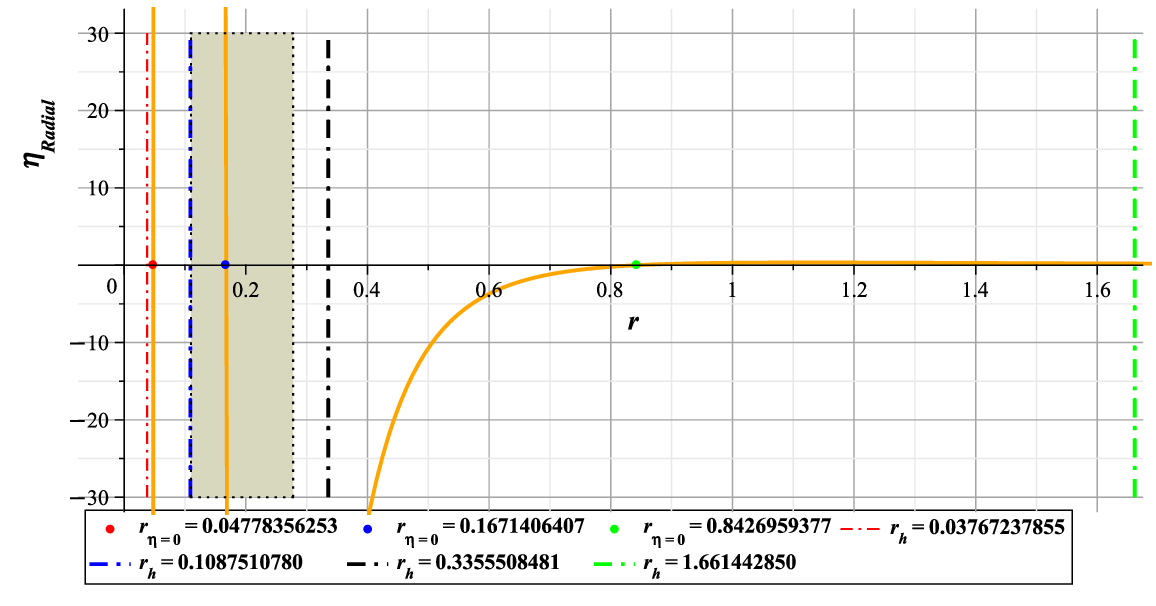}
    \hfill
\caption{ $\eta_{Radial}$  function with $\alpha_{3} = 6.38736 \times 10^{-7},\alpha_{2} = -0.004, b = 100, M = 1$ for 18a: $q=0.373$, 18b: $q=0.49$, 18c: $q=0.75$  in VI-Horizon structure of  NED Black hole.}
    \label{18}
\end{figure}
\begin{itemize}
\item \textbf{Schwarzschild-like region:}\\ Strictly speaking, the existence of this zero crossing distinguishes the solution from the Schwarzschild spacetime, since the corresponding tidal-force component changes sign once. However, an important point must be emphasized: as the particle approaches this zero point, the intensity of the force variation becomes so extreme that, for example, in Fig.18a with the chosen parameters, the magnitude of the radial tidal force component near the innermost zero point reaches approximately $5.6 \times 10^6$. Relative to the remainder of the trajectory, this value is sufficiently large that the corresponding region is unlikely to be physically accessible for an extended body. Under this assumption, the tidal force in the region under study will not experience any further sign change.

\item \textbf{New Behavioral Region region:}\\  A similar argument applies in this regime. The tidal force again becomes extremely large near the innermost zero crossing, making this region physically inaccessible for any realistic extended body. In contrast, the remaining two zero crossings occur sufficiently far from the innermost horizon and are associated with finite tidal forces, making them accessible along the trajectories considered here. In this region, the qualitative behavior pattern agrees with that observed for the three-horizon solution: namely, the infalling body experiences a novel behavior in which it senses two sign changes in the force structure (from stretching to compressing and vice versa), as illustrated in Fig.18b.

\item \textbf{Protected region:} In this regime the qualitative behavior changes. The two innermost zero crossings lie inside the classically forbidden region and therefore cannot be reached by particles following the trajectories considered in this work. Consequently, the dual protective mechanism discussed previously remains fully operative in this regime, Fig.18c.
\end{itemize}  
In the super-extrem charges regime, the four-horizon solution exhibits the same qualitative behavior as the corresponding three-horizon solution (Fig.~\ref{19}).
The combination of finite tidal forces outside the event horizon and the expansion of the classically forbidden region further reinforces the protective behavior identified in the three-horizon solution. Within the framework of the present model, these features suggest that nonlinear electromagnetic effects can substantially modify the dynamics of infalling particles in the vicinity of the black hole.   
\begin{figure}[htbp]
    \centering
    \includegraphics[width=1.01\linewidth, height=4.8cm ]{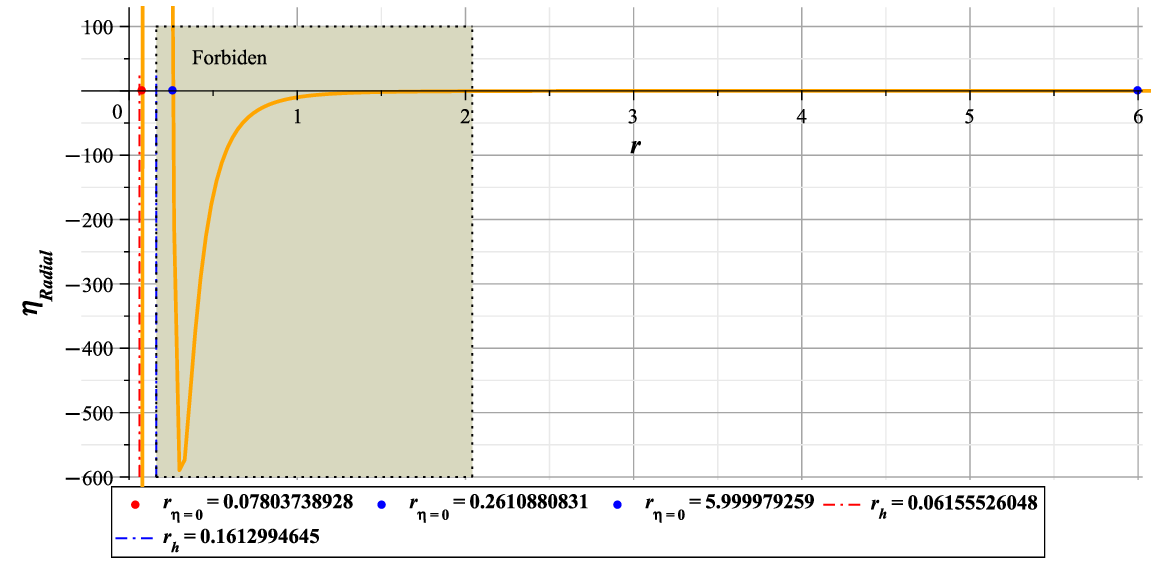}
    \hfill
    \includegraphics[width=1.01\linewidth, height=4.8cm ]{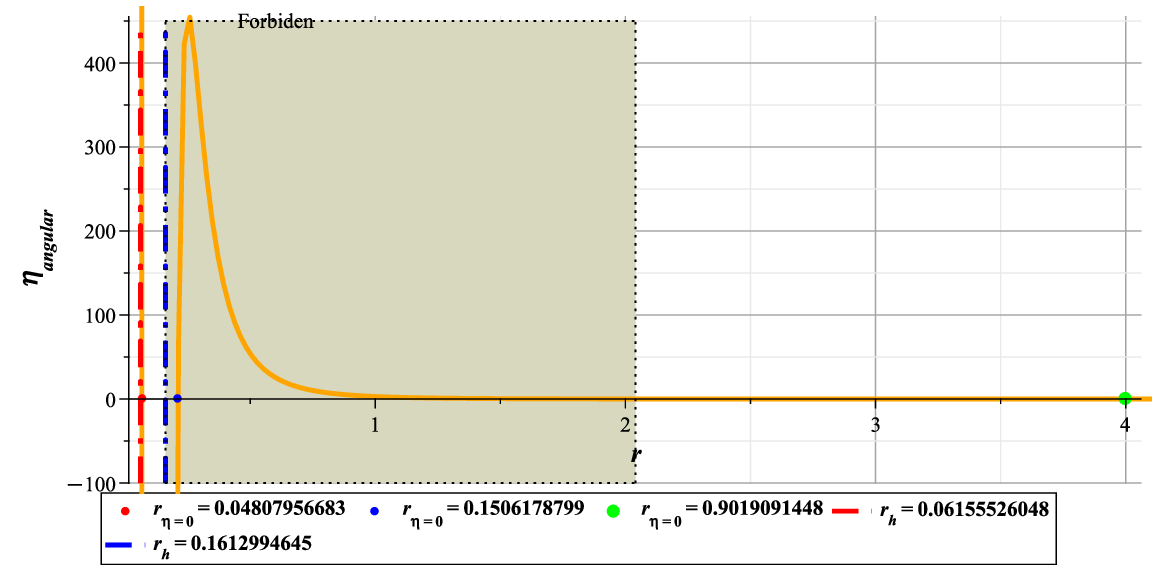}
    \hfill
         \caption{19a: $\eta_{radial}$ and 19b: $\eta_{angular}$   function with $\alpha_{3} = 6.38736 \times 10^{-7},\alpha_{2} = -0.004, b = 100, M = 1,q=2 $ for VI-Horizon structure of  NED Black hole.}
    \label{19}
\end{figure}  
Finally, in Fig. ~\ref{20}, we can have a comprehensive look at the behavior of the tidal force components in the 4-horizon structure for different values of $ q $
\begin{figure}[htbp]
    \centering
    \includegraphics[width=1.01\linewidth, height=4.8cm ]{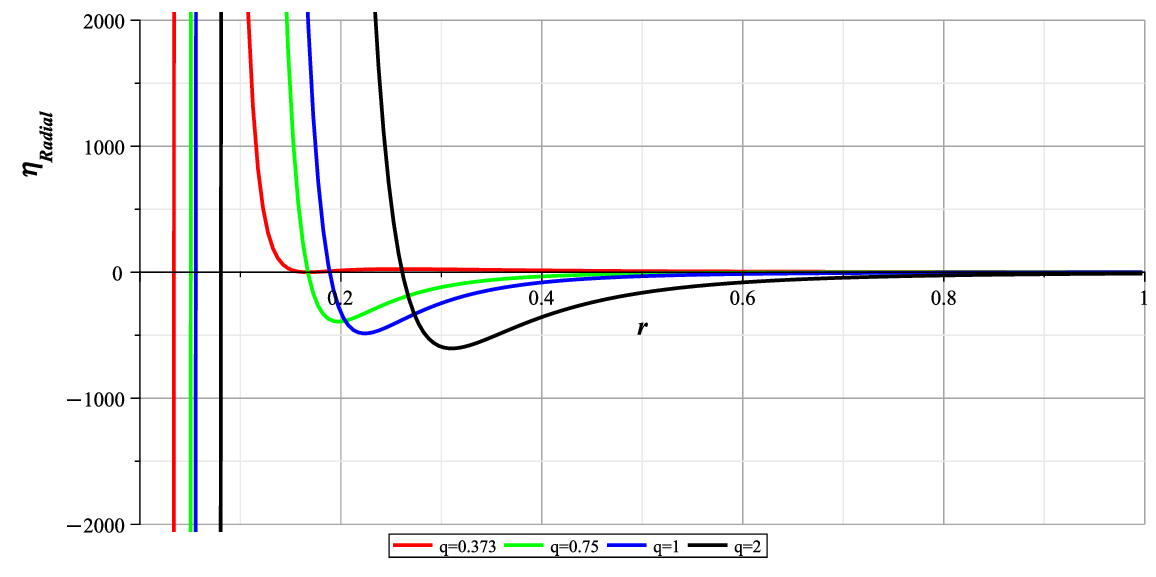}
    \hfill
    \includegraphics[width=1.01\linewidth, height=4.8cm ]{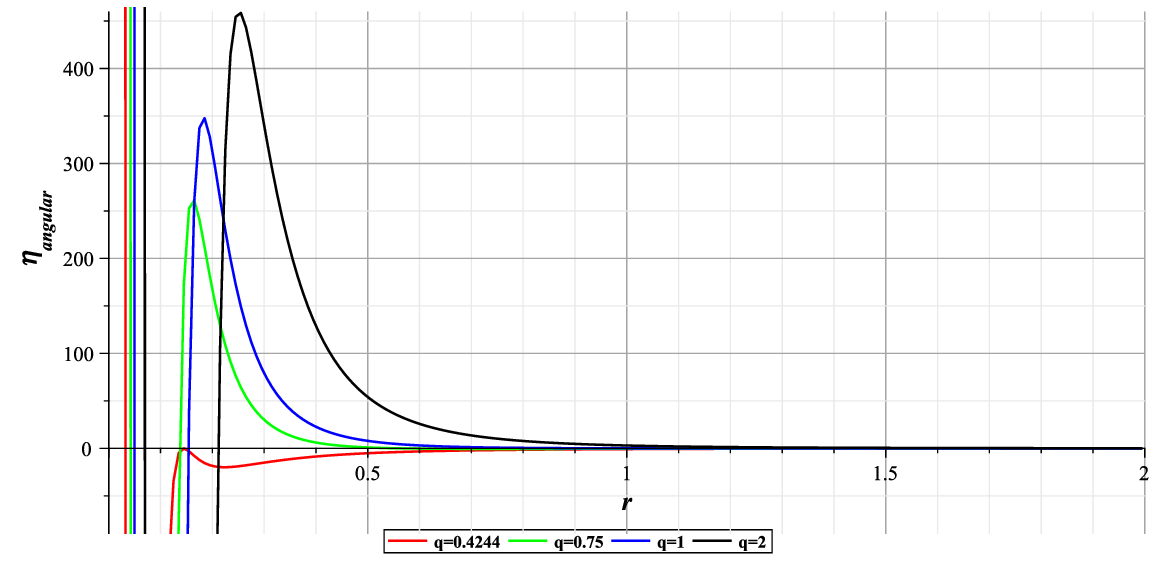}
    \hfill
         \caption{20a: $\eta_{radial}$ and 20b: $\eta_{angular}$   function set of behaviors for different charges in III-Horizon structure of  NED Black hole.}
    \label{20}
\end{figure}    
 \subsection{ Geodesic Deviation Equations}   
Given the general form of the metric function (\ref{eq:metric}) and energy, by substituting \ref{eq:rad_force} and \ref{eq:ang_force} into \ref{eq:deviation} and solving the differential equations, the geodesic deviation for the horizon form of the 4 NED black holes will be obtained as follows:  
\begin{equation*}
\begin{split}
&\kappa \! \left(r \right)=-\frac{2 M}{b}+\frac{q^{2}}{b^{2}}+\frac{2 \alpha_{2} q^{4}}{5 b^{6}}+\frac{q^{6} \left(16 \alpha_{2}^{2}-4 \alpha_{3}\right)}{9 b^{10}}+\\ &\frac{2 M}{r}-\frac{q^{2}}{r^{2}}-\frac{2 \alpha_{2} q^{4}}{5 r^{6}}-\frac{q^{6} \left(16 \alpha_{2}^{2}-4 \alpha_{3}\right)}{9 r^{10}}\\,
\end{split}
\end{equation*}
\begin{equation}
\begin{split}
&\xi^{\hat{1}}(r) = \sqrt{\kappa(r)}\times[\\ & \frac{45 r^{11} \dot{\xi}^{\hat{1}}(b)}{100 \left(-4 \alpha_{2}^{2}+\alpha_{3}\right) q^{6}-54 q^{4} \alpha_{2} r^{4}-45 q^{2} r^{8}+45 M \,r^{9}} +\\ & \left( \frac{M}{r^{2}}-\frac{q^{2}}{r^{3}}-\frac{6 q^{4} \alpha_{2}}{5 r^{7}}-\frac{20 \left(4 \alpha_{2}^{2}-\alpha_{3}\right) q^{6}}{9 r^{11}} \right)\\ & \times \xi^{\hat{1}}(b) \int \frac{dr}{\kappa(r)^{3/2}}] \\,
\end{split}
 \end{equation} 
 \begin{equation}
\xi^{\hat{i}}=r \left(\frac{\xi^{\hat{i}}(b)}{b}-b \dot{\xi}^{\hat{i}}(b) \int \frac{1}{\sqrt{\kappa(r)}} dr \right).
\end{equation}  
\section{Discussion and Synthesis of Results}

In this section, to ensure coherence, deepen the analyses, and achieve a comprehensive understanding of the emergent phenomena, we examine the obtained results collectively within a logical sequence. We considered a black hole metric structure whose action incorporates a nonlinear electrodynamics (NED) field and exhibits a multi-horizon configuration in specific regions of the parameter space. Our primary objective was to study the tidal forces experienced by a neutral test particle released from rest at a large distance ($b\gg r_h$) and falling radially toward the black hole. As previously shown for the RN spacetime, the electromagnetic field can introduce zero crossings in the tidal-force components, producing transitions between stretching and compressive tidal regimes.

Since the NED corrections introduce higher-order inverse powers of the radial coordinate (e.g., $r^{-6}$ for the three-horizon solution and $r^{-10}$ for the four-horizon solution), additional roots of both the radial equation of motion and the tidal-force components naturally arise. However, our studies revealed that this increase is not merely a linear algebraic consequence; rather, the integration of this mathematical structure with physical concepts leads to complex and novel phenomena, which we elaborate on below:

\textbf{1. Multiple Tidal-Force Transitions:}\\
As expected, the additional zero crossings of the tidal-force components cause the infalling particle to experience multiple transitions between stretching and compressive tidal regimes instead of the single transition characteristic of the R-N spacetime. These multiple sign reversals reflect the richer tidal structure associated with the multi-horizon NED geometry.

\textbf{2. Dynamical Forbidden Region and Kinematic Inaccessibility of the Singularity:}\\
A deeper finding of this study is the emergence of a classically forbidden region bounded by successive bounce-back points. Within the framework of the trajectories considered in this work, this forbidden region has two important physical consequences:\\
First, for particles released from rest under the initial conditions considered here, the forbidden region prevents access to the central singularity. Unlike regular black hole solutions such as the Bardeen or Hayward geometries, where the central singularity is removed, the singularity remains present in the current solution. Nevertheless, for the class of trajectories studied here it becomes \textit{kinematically inaccessible} because the particle reverses its motion before reaching it. Phenomenologically, this phenomenon mimics the behavior of regular black holes and suggests that it is possible that the coupling of certain nonlinear gauge fields (such as this one) with gravity could prevent physical access to the singularity without requiring its removal.\\
Second, the forbidden region modifies the capture of particles by the black hole. As the charge increases toward the super-extremal form considered in this work, the forbidden region expands and eventually extends outside the event horizon. Consequently, neutral particles released from rest at sufficiently large distances reverse their motion before crossing the event horizon. Within the present model and for the class of trajectories considered here, this behavior suppresses the classical capture of neutral particles released from rest.

\textbf{3. Avoidance of Divergent Tidal Forces:}\\
In classical black hole spacetimes such as the Schwarzschild and R-N solutions, the divergence of the tidal-force components near the singularity is responsible for the onset of spaghettification and the eventual destruction of extended bodies. For the class of trajectories considered in this work, the classically forbidden region prevents the particle from entering the spacetime region where the tidal forces become divergent. Consequently, the particle reverses its motion before encountering the divergent stretching or compressive tidal forces characteristic of the Schwarzschild and RN spacetimes. This feature suggests that some types of NED fields may be able to prevent the onset of divergent tidal forces along paths such as those considered here, thereby providing conditions under which the structural integrity of objects and particles near the black hole horizon is preserved.

\textbf{4. Sequential Emergence of Critical Transitions:}\\
the most  noteworthy feature in this work is the sequence in which these phenomena emerge as the black hole charge increases. According to Tables~\ref{tab:3horizon} and \ref{tab:4horizon}, the three-horizon solution exhibits successive critical values corresponding to the appearance of the radial tidal-force zero crossings ($q=0.547$), the angular tidal-force zero crossings ($q=0.624$), the additional horizons ($q=0.75$), and finally the bounce-back points ($q=0.8223$). The same qualitative ordering is observed for the four-horizon solution, although the corresponding critical charge values differ: $q_{\mathrm{rad}}=0.374$, $q_{\mathrm{ang}}=0.4245$, $q_H=0.54$, and $q_{\mathrm{TP}}=0.561$.)
\begin{table*}[htbp]
\centering
\caption{Variations of different parameters with respect to $q$ for three-horizon structure with $\alpha_{2} = -0.004, b = 100, M = 1$ in NED black hole.}
\label{tab:3horizon}
\resizebox{\textwidth}{!}{%
\begin{tabular}{ccccccccccccc} 
\hline\hline
$q$ & $r_h^{(1)}$ & $r_h^{(2)}$ & $r_h^{(3)}$ & $r_{\text{stop}}^{(1)}$ & $r_{\text{stop}}^{(2)}$ & $r_{\text{stop}}^{(3)}$ & $r[\eta_{\text{radial}}=0]$ & $r[\eta_{\text{radial}}=0]$ & $\eta_{\text{radial}}^{\text{Min}}$ & $\eta_{\text{radial}}^{\text{Max}}$ & $r[\eta_{\text{angular}}=0]$ & $r[\eta_{\text{angular}}=0]$ \\
\hline
0.547 & 1.83714 & 0.00000 & 0.00000 & 100.00000 & 0.00000 & 0.00000 & 0.00000 & 0.00000 & 0.08724 & 2.57261 & 0.00000 & 0.00000 \\
0.624 & 1.78144 & 0.00000 & 0.00000 & 100.00000 & 0.00000 & 0.00000 & 0.31069 & 0.55774 & -8.65200 & 1.10051 & 0.30274 & 0.31979 \\
0.650 & 1.75995 & 0.00000 & 0.00000 & 100.00000 & 0.00000 & 0.00000 & 0.31031 & 0.61243 & -11.34536 & 0.85370 & 0.27703 & 0.38245 \\
0.749 & 1.66262 & 0.00000 & 0.00000 & 100.00000 & 0.00000 & 0.00000 & 0.31684 & 0.83038 & -20.21003 & 0.35834 & 0.26795 & 0.54372 \\
0.750 & 0.26423 & 0.27195 & 1.66149 & 100.00000 & 0.00000 & 0.00000 & 0.31694 & 0.83269 & -20.28898 & 0.35545 & 0.26797 & 0.54533 \\
0.800 & 0.23051 & 0.37218 & 1.60008 & 100.00000 & 0.00000 & 0.00000 & 0.32230 & 0.95161 & -23.99595 & 0.24037 & 0.26999 & 0.62730 \\
0.822 & 0.22755 & 0.40826 & 1.56916 & 100.00000 & 0.27058 & 0.27212 & 0.32490 & 1.00679 & -25.50703 & 0.20356 & 0.27134 & 0.66495 \\
0.850 & 0.22566 & 0.45504 & 1.52693 & 100.00000 & 0.24397 & 0.32476 & 0.32824 & 1.07724 & -27.27294 & 0.16665 & 0.27328 & 0.71279 \\
0.920 & 0.22503 & 0.59671 & 1.39231 & 100.00000 & 0.23462 & 0.40311 & 0.33705 & 1.26490 & -31.24717 & 0.10344 & 0.27902 & 0.83948 \\
0.980 & 0.22674 & 0.79178 & 1.20077 & 100.00000 & 0.23380 & 0.46683 & 0.34485 & 1.43697 & -34.17638 & 0.07072 & 0.28449 & 0.95508 \\
1.000 & 0.22756 & 0.95626 & 1.03718 & 100.00000 & 0.23408 & 0.48830 & 0.34747 & 1.49665 & -35.06844 & 0.06263 & 0.28638 & 0.99510 \\
2.000 & 0.29429 & 0.00000 & 0.00000 & 100.00000 & 0.29429 & 2.04005 & 0.46953 & 5.99979 & -55.07075 & 0.00098 & 0.38169 & 3.99970 \\
\hline\hline
\end{tabular}%
}
\end{table*}

\begin{table*}[htbp]
\centering
\caption{Variations of different parameters with respect to $q$ for four-horizon structure with $\alpha_{3} = 6.38736 \times 10^{-7}, \alpha_{2} = -0.004, b = 100, M = 1$ in NED black hole.}
\label{tab:4horizon}
\resizebox{\textwidth}{!}{%
\begin{tabular}{cccccccccccccc} 
\hline\hline
$q$ & $r_h^{(1)}$ & $r_h^{(2)}$ & $r_h^{(3)}$ & $r_h^{(4)}$ & $r_{\text{stop}}^{(1)}$ & $r_{\text{stop}}^{(2)}$ & $r_{\text{stop}}^{(3)}$ & $r_{\text{stop}}^{(4)}$ & $r[\eta_{\text{radial}}=0]$ & $r[\eta_{\text{radial}}=0]$ & $r[\eta_{\text{radial}}=0]$ & $r[\eta_{\text{angular}}=0]$ & $r[\eta_{\text{angular}}=0]$ \\
\hline
0.373 & 0.02653 & 1.92783 & 0.00000 & 0.00000 & 100.00000 & 0.02653 & 0.00000 & 0.00000 & 0.03369 & 0.00000 & 0.00000 & 0.00000 & 0.00000 \\
0.374 & 0.02657 & 1.92743 & 0.00000 & 0.00000 & 100.00000 & 0.02656 & 0.00000 & 0.00000 & 0.03374 & 0.16106 & 0.17403 & 0.00000 & 0.00000 \\
0.424 & 0.02831 & 1.90548 & 0.00000 & 0.00000 & 100.00000 & 0.02831 & 0.00000 & 0.00000 & 0.03594 & 0.14404 & 0.25785 & 0.03212 & 0.00000 \\
0.425 & 0.02831 & 1.90543 & 0.00000 & 0.00000 & 100.00000 & 0.02831 & 0.00000 & 0.00000 & 0.03594 & 0.14403 & 0.25799 & 0.03213 & 0.14256 \\
0.490 & 0.03043 & 1.87172 & 0.00000 & 0.00000 & 100.00000 & 0.03043 & 0.00000 & 0.00000 & 0.03862 & 0.14558 & 0.35398 & 0.03452 & 0.12409 \\
0.530 & 0.03165 & 1.84800 & 0.00000 & 0.00000 & 100.00000 & 0.03165 & 0.00000 & 0.00000 & 0.04017 & 0.14823 & 0.41696 & 0.03590 & 0.12454 \\
0.540 & 0.03195 & 0.11849 & 0.13289 & 1.84167 & 100.00000 & 0.03195 & 0.00000 & 0.00000 & 0.04054 & 0.14898 & 0.43335 & 0.03624 & 0.12486 \\
0.560 & 0.03254 & 0.11048 & 0.15542 & 1.82849 & 100.00000 & 0.03254 & 0.00000 & 0.00000 & 0.04129 & 0.15054 & 0.46692 & 0.03691 & 0.12566 \\
0.570 & 0.03283 & 0.10896 & 0.16444 & 1.82165 & 100.00000 & 0.03283 & 0.11561 & 0.14185 & 0.04165 & 0.15136 & 0.48412 & 0.03724 & 0.12612 \\
0.650 & 0.03506 & 0.10628 & 0.23381 & 1.75994 & 100.00000 & 0.03506 & 0.10816 & 0.20331 & 0.04448 & 0.15822 & 0.63187 & 0.03976 & 0.13062 \\
0.750 & 0.03767 & 0.10875 & 0.33555 & 1.66144 & 100.00000 & 0.03767 & 0.10984 & 0.27777 & 0.04778 & 0.16714 & 0.84270 & 0.04272 & 0.13714 \\
0.800 & 0.03891 & 0.11063 & 0.39782 & 1.60001 & 100.00000 & 0.03891 & 0.11155 & 0.31779 & 0.04935 & 0.17159 & 0.95919 & 0.04412 & 0.14051 \\
0.850 & 0.04011 & 0.11269 & 0.47162 & 1.52680 & 100.00000 & 0.04011 & 0.11349 & 0.36006 & 0.05087 & 0.17599 & 1.08311 & 0.04548 & 0.14390 \\
0.950 & 0.04241 & 0.11707 & 0.68681 & 1.31232 & 100.00000 & 0.04241 & 0.11772 & 0.45172 & 0.05378 & 0.18463 & 1.35334 & 0.04808 & 0.15062 \\
1.000 & 0.04351 & 0.11932 & 0.98702 & 1.01234 & 100.00000 & 0.04351 & 0.11992 & 0.50123 & 0.05518 & 0.18885 & 1.49967 & 0.04933 & 0.15393 \\
2.000 & 0.06156 & 0.16130 & 0.00000 & 0.00000 & 100.00000 & 0.06155 & 0.16159 & 2.04074 & 0.07804 & 0.26109 & 5.99998 & 0.06977 & 0.21157 \\
\hline\hline
\end{tabular}%
}
\end{table*}
In both cases, the critical charges satisfy the hierarchy:
\begin{equation}
q_{\text{rad}} < q_{\text{ang}} < q_H < q_{\text{TP}}
\end{equation}

This ordering suggests that the spacetime undergoes a sequence of qualitatively distinct transitions as the black hole charge increases. In particular, the tidal-force structure responds to increasing charge before the additional horizons appear, suggesting that tidal observables provide an earlier indicator of changes in the horizon structure. The sudden appearance of new roots at well-defined critical charges is qualitatively reminiscent of saddle-node bifurcations encountered in nonlinear dynamical systems. A rigorous identification of these transitions as saddle-node bifurcations would require an independent stability analysis and therefore lies beyond the scope of the present work. Nevertheless, the systematic appearance of new roots in the tidal-force components, horizon structure, and radial motion at successive critical charge values strongly suggests an underlying bifurcation-like behavior. 
  
\section{Conclusion}

In this work, we investigated the radial motion and tidal forces experienced by neutral test particles in multi-horizon black hole solutions arising from Einstein gravity coupled to nonlinear electrodynamics (NED). Particular attention was devoted to the three- and four-horizon configurations and to the influence of the black hole charge on the causal structure, particle dynamics, and tidal behavior.

Our analysis shows that the nonlinear electromagnetic corrections introduce qualitatively new features that are absent in the Schwarzschild and Reissner--Nordström spacetimes. In addition to multiple zero crossings of the radial and angular tidal-force components, the radial equation of motion develops classically forbidden regions bounded by bounce-back points. As a consequence, particles following the trajectories considered in this work reverse their motion before reaching the regions where the tidal forces become divergent. In the super-extremal regime admitted by these NED solutions, the forbidden region may extend beyond the event horizon, preventing particles released from rest at sufficiently large distances from crossing the horizon.

Another noteworthy result is the existence of a well-defined sequence of critical charge values associated with the appearance of tidal-force zero crossings, additional horizons, and bounce-back points. For both the three- and four-horizon configurations, these transitions satisfy the hierarchical ordering
\begin{equation*}
q_{\text{rad}} < q_{\text{ang}} < q_H < q_{\text{TP}}
\end{equation*}
indicating that changes in the tidal-force structure precede the corresponding modifications of the horizon configuration. This systematic ordering suggests the existence of an underlying bifurcation-like behavior in the dynamical evolution of the spacetime as the charge increases.

Overall, the present results demonstrate that nonlinear electrodynamics can substantially modify the classical dynamics of neutral particles in black hole spacetimes without removing the central singularity. Instead, the nonlinear electromagnetic field alters the accessible region of spacetime through the formation of classically forbidden regions and modifies the corresponding tidal evolution. These findings provide further insight into the rich phenomenology of multi-horizon NED black holes and motivate future investigations of their stability, observational signatures, and extensions to more general classes of trajectories.

\end{document}